\documentclass{aa}  
\usepackage{natbib}
\usepackage{graphicx}
\usepackage[T1]{fontenc}

\usepackage{caption}
\usepackage{subcaption}
\usepackage{natbib}
\usepackage{float}
\usepackage{longtable}
\usepackage{multicol}
\usepackage{multirow}
\usepackage{gensymb}
\usepackage{placeins}
\usepackage{txfonts}
\usepackage{graphicx}	
\usepackage{amsmath}	
\usepackage[english]{babel} 
\usepackage{blindtext}
\usepackage{wrapfig}
\usepackage{textcomp}
\usepackage{lineno}
\usepackage{gensymb}
\usepackage[x11names]{xcolor}

\usepackage[]{hyperref}

\usepackage{newtxtext,newtxmath}
\defcitealias{kk20}{KK20}
\defcitealias{pm23}{PM23}
\defcitealias{l22}{L22}
\begin{document}

   \title{Post-processing of galaxies due to major cluster mergers}

   \subtitle{I. hints from galaxy colours and morphologies }

   \author{K. Kelkar
          \inst{1,2}\thanks{E-mail: kkshitija.astro@gmail.com}
          \and
          Y. L. Jaff\'e\inst{1,2}
          \and
          A. C. C. Louren\c{c}o\inst{2,4}
          \and
          D. P\'erez-Mill\'an\inst{3}\fnmsep
          \and
          J. Fritz\inst{3}
          \and
          B. Vulcani\inst{5} 
          \and
          J. P. Crossett\inst{1,2}
          \and
          B. Poggianti \inst{5}
          \and
          A. Moretti\inst{5}
          }

   \institute{Universidad Técnica Federico Santa María, Av. España 1680, Valpara\'iso, Chile\\
            \and
            Instituto de F\'isica y Astronomia, Universidad de Valpara\'iso, 1111 Gran Breta\~{n}a, Valpara\'iso, Chile\\
        \and
            Instituto de Radioastronomia y Astrofisica, UNAM, Campus Morelia, Michoac\'an, C.P. 58089, Mexico\\   
        \and
            European Southern Observatory (ESO), Alonso de Cordova 3107, Santiago, Chile \\
        \and
            INAF-Osservatorio astronomico di Padova, Vicolo Osservatorio 5, IT-35122 Padova, Italy\\}

   \date{Received August 04, 2023; accepted September 22, 2023}

 
  \abstract {Galaxy clusters, which underwent a recent ($\leq$3 Gyr) major merger, offer a harsher environment due to the global hydrodynamical disturbance and the merger-shock heated ICM. However, the aftermath of such extreme cluster interactions on the member galaxy properties is not very well constrained. We explore the integrated star formation properties of galaxies through galaxy colours, as well as morphology buildup in three nearby ($0.04<z<0.07$) young ($\sim$0.6-1 Gyr) post-merger clusters -- A3667, A3376 \& A168-- and 7 relaxed clusters, to disentangle merger-induced post-processing signatures from the expected effects due to high-density cluster environments. Exploiting the optical spectroscopy \& photometry from the OmegaWINGS survey, we find that post-merger clusters are evolved systems demonstrating uniform spiral fractions, uniform fraction of blue galaxies and constant scatter in the colour--magnitude relations, a regularity that is absent in dynamically relaxed clusters. While no clear merger-induced signatures were revealed in the global colours of galaxies, we conclude that different global star formation histories of dynamically relaxed clusters lead to considerable scatter in galaxy properties, resulting in the pre-merger cluster environment to potentially contaminate any merger-induced signal in galaxy properties. We discover red spirals to be common to both post-merger and relaxed clusters while post-merger clusters appear to host a non-negligible population of blue early-type galaxies. We propose that while such merging cluster systems absorb extra cosmic web populations hitherto not part of the original merging subclusters, a $\sim$ 1 Gyr timescale is possibly insufficient to result in changes in global colours and morphologies of galaxies. }


   \maketitle
%
\section{Introduction}
Hierarchical growth of cosmic structures dictates that over time galaxies become part of increasingly larger systems like groups, clusters \& super-clusters which are connected through cosmic filaments \citep{press74,fakhouri10b}. We know thus far that such dense environments influence the structure and star formation in galaxies differently than if they were left to evolve in isolation. Indeed, galaxies with different morphologies are found to preferentially live in different broad areas of a cluster \citep[ e.g. spiral galaxies avoid dense cluster cores;][]{dressler1980, poggianti09,vulcani23}. Moreover, galaxies are also found to stop their star formation in such environments and change their structure along this journey \citep{vulcani10,prescott11,petropoulou12,haines13,grootes17,kk17,kelkar19,lopes16,lopes17,burchett18,matharu19,li19}, thereby strongly advocating for environmentally driven gas removal processes like ram-pressure stripping \citep[see reviews by][]{cortese21, boselli22} \& galaxy starvation \citep{larson80}. Current wide-field galaxy surveys are now revealing the complex continuous nature of such large-scale structures around galaxies which can no longer be encapsulated in simple `field/isolated' and `cluster' definitions of galaxy environments. Consequently, new studies are reporting that galaxies get `pre-processed' in intermediate environments like groups \& filaments that can contribute to their quenching prior to entering a massive cluster \citep[e.g.][]{fujita04,mcgee09,dressler13,bahe13,haines15,jaffe16,kraljic18,vulcani19,kuchner22}. 

The channels of growth of galaxy clusters however are not just limited to infalling galaxies/cosmic filaments/group accretion but also cumulative explosive megaparsec-scale events like cluster-cluster mergers. These powerful interactions are rare events, and capable of disturbing the cluster as a whole by heating the intracluster medium (ICM), thus affecting the morphology of the system \citep[see e.g.][]{caglar17,botteon18,caglar18}. These effects have become beacons for detecting such merging cluster systems through X-ray emission from the hot ICM, and non-thermal processes such as diffuse synchrotron radio emission in the form of halos \citep[located centrally in a cluster with a history of major interaction, e.g.][]{kale22} and relics \citep[detected towards cluster peripheries; See also review by ][]{vweeren19}. While halos generally tell us about ICM turbulence and the dynamic history of clusters, radio relics arise when substantial amounts of kinetic energy are released during cluster merging events (also known as the merger shock). Hence, the presence of such shockfront, observed at times in X-ray emission as well \citep{akamatsu13,bourdin13,sarazin16, akamatsu17-b,gu19}, indicates a very recent ($\sim$ 3 Gyr) violent dynamic major interaction in between galaxy clusters. With only $\sim$70 radio relics (and $<$20 twin or `double' radio relics)  detected till date \citep{golovich19b, knowles22}, their detectability depends much on the merger geometry \citep[e.g. plane of the sky mergers,][]{golovich19a}, the mass ratio of the participating clusters, age of the merger, turbulence decay timescale, and synchrotron life-time of decaying electrons \citep{brunetti14}. Furthermore, simulations demonstrate that while the shock propagates for $\sim$2 Gyr since the collision, the injected turbulence in the ICM sustains for at least $\sim$4 Gyr \citep{paul11}. This makes cluster mergers exotic systems with an extreme environment that can potentially leave unique signatures on the galaxies or \lq{post-process}\rq\hspace{0.05cm} the galaxies within.  

To date, such post-processing signatures have been traced throughout the spread of cluster interactions and the associated growing large-scale structure. Recent comprehensive studies show that generally disturbed clusters showing ongoing interaction (like group accretion, minor merging, etc) could result in possibly mild enhancement in the fraction of ram-pressure stripped galaxies (e.g. \citet{l22}), generally increased star formation activity \citep[e.g][]{cohen14}, enhanced star formation in barred galaxies \citep[e.g.][]{yoon20}, presence of younger AGN population \citep[e.g.][]{bilton20}. Alternatively, some studies show that interacting clusters present no unique environment any different than relaxed clusters \citep[e.g.][]{shim11,kleiner14}. Major merging clusters that have recently undergone core passage, reportedly show evidence of both enhanced star formation \citep[][]{stroe15b,stroe15-c} and quenching \citep[e.g.][]{pranger14} or a potentially net zero effect on star formation activity \citep[e.g.][]{chung09,rawle14}. Moreover, the wider large-scale structure surrounding these major mergers are revealing heightened star-forming activity well beyond $R_{200}$ \citep[][]{stroe17,stroe21} or enhanced ram-pressure stripping \citep{franco23}. As a consequence it is difficult to distinguish the merger-induced post-processing signatures without the knowledge of cluster galaxy properties prior to the merging event, thus highlighting the need to account for the overall \lq{cumulative}\rq\hspace{0.05cm}processing of galaxies \citep[][ hereafter KK20]{mansheim17-b,kk20}.

In this paper, we attempt to constrain such cumulative environmental signatures i.e. the combined effect of quenching and morphological transformation of galaxies in dense environment, and segregate post-processing effects on the integrated star formation of galaxies and galaxy morphologies. We present a population study of galaxy properties between three cluster mergers with radio relics Abell 3667, Abell 3376 \& Abell 168 -- all of whom experienced a major merger over similar time since core collision (TSC) of $\sim$0.6-1 Gyr -- in comparison to relaxed clusters within similar redshift range of $0.04<z<0.07$, and using the same coverage-matched dataset. The paper is organised as follows - Section \ref{data} details the cluster merger sample and galaxy sample we have used in this analysis. Thereafter we present our analysis and results starting from galaxy distributions and morphology fractions in post-merger clusters (Section \ref{sec:morph}), followed by rest-frame galaxy colour--magnitude relations for post-merger and relaxed clusters, and consequent integrated star formation in post-merger clusters inferred through red and blue fractions (Section \ref{sec:gcolors}). Finally, we bring all the results together to tell us the story of how different galaxy populations in cluster mergers really are in comparison to relaxed clusters without any interactions (Section \ref{sec:disc}) with Section \ref{sec:conc} summarising our key results. Throughout this paper, we use the standard $\Lambda$CDM cosmology ($h_0$=0.7, $\Omega_\Lambda$=0.7 and  $\Omega_{\rm m}$=0.3), and \citet{chabrier03} initial mass function.

\section{Data}
\label{data}

\subsection{Cluster sample}
\label{data-cl}

The choice of our cluster sample is motivated by understanding galaxy evolution in two extremely different environments-- post-mergers \footnote{Throughout this work, \lq post-mergers\rq are defined as systems which have undergone a major merging event, which includes a recent core passage and are currently in the process of reaching maximum separation before coalescing into a single system. They display at least one radio relic, with confirmation of TSC through radio and if available X-ray studies. However, except for A3376 (See Section \ref{data-cl}),  we have not yet ascertained the precise merging dynamics of the other clusters using simulations.} and dynamically relaxed non-merging clusters -- controlling for dynamic specifics as much as possible (e.g. similar TSC of $\sim$ 0.6-1 Gyr in case of merging clusters). Moreover, the narrow $z-$range of $0.04<z_{cl}<0.065$ makes this the lowest redshift uniform cluster sample spanning the most extreme dynamical stages.

We build our homogeneous cluster sample from the OmegaWINGS survey \citep{gull15,moretti17} which is a photometric survey of 57 galaxy clusters with a spatial coverage of $\sim1$ sq degree ($\sim$2.5 virial radii ). Originally based on the WIde-field Nearby Galaxy-cluster Survey \citep[WINGS;][]{fasano06,moretti14} comprising 76 clusters, OmegaWINGS provides photometric and imaging data in the $U-$ \citep{omizzolo14,donofrio20}, $B-$, $V-$ bands using the $OmegaCAM/VST$, and spectroscopy of subsample of 33 out of 57 clusters using $AAOmega$ spectrograph at the Anglo Australian Telescope (AAT) \citep{cava09,moretti17}. Global cluster properties like the mean cluster redshift $z_{\rm cl}$, and the cluster velocity dispersion $\sigma_{\rm cl}$ were iteratively determined through $3\sigma$ clipping using the biweight robust location and scale estimators \citep{beers90}. Galaxies were assigned cluster memberships if they lie within $3\sigma_{\rm cl}$ from $z_{\rm cl}$ \citep{moretti17} while the physical radius $R_{200}$ was provided by \citet{biviano17}. 

\begin{figure}
	
	\includegraphics[width=\columnwidth]{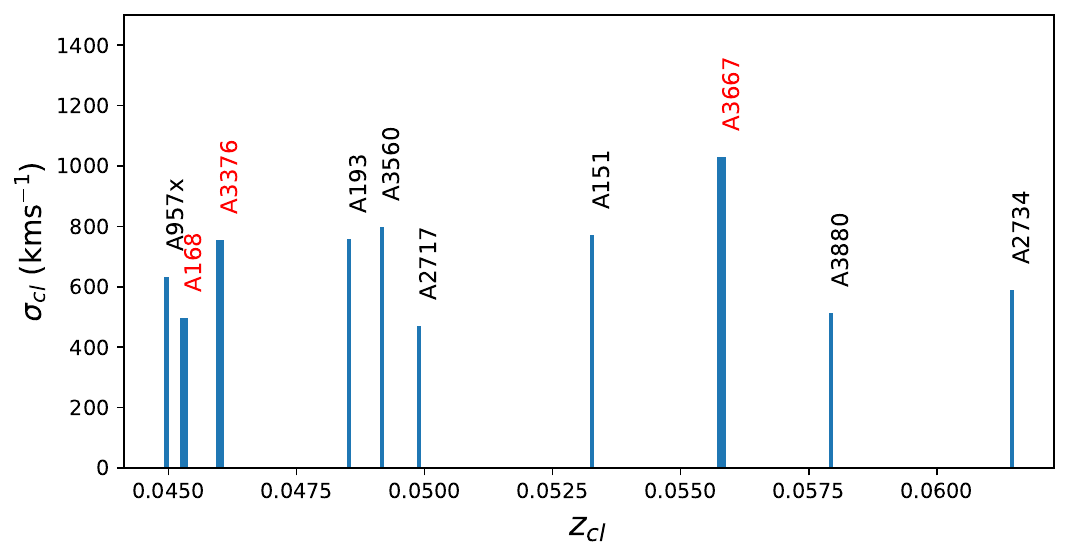}
    \caption{Summary of post-merger (thick bars annotated in red) and relaxed control cluster sample (thin bars annotated in black) as a function of cluster redshift and velocity dispersion.}
    \label{fig:sigz}
\end{figure}

We target three extreme post-merging galaxy clusters from OmegaWINGS - Abell 3667 (A3667), Abell 3376 (A3376; pilot analysis by \citetalias{kk20}) and Abell 168 (A168). Complementary to this target cluster merger sample, we also define an ancillary control sample of clusters to isolate merger-induced post-processing signatures in member galaxies. We utilise the control sample originally presented by \citetalias{kk20}, where they visually inspected the X-ray images from $XMM-Newton$ telescope to look for signatures of interaction through X-ray surface brightness disturbances. We also include clusters that are identified to be dynamically relaxed through a systematic analysis of dynamical stages of WINGS and OmegaWINGS clusters performed by \citet[][hereafter L22]{l22}, using optical information from WINGS/OmegaWINGS and X-ray data from $Chandra$ and/or $XMM-Newton$ telescopes. Specifically, they use the positions of the BCG(s) with respect to the X-ray peak, the morphology of X-ray surface brightness distribution (e.g. concentrated, asymmetric, presence of secondary peaks), and detection of radio relics from literature  to identify a dynamical state sequence for the galaxy clusters. The final dynamic states classifications were thus defined as `pre-merger' ($1$), `relaxed' ($2$), `mildly interacting' ($3$), `interacting' ($4$) \& `post-merger' ($5$). This classification of clusters' dynamical states also recovers the control cluster sample of \citetalias{kk20}, except for one cluster. Thus, our final control cluster sample in this work consists of seven OmegaWINGS clusters ($0.04<z_{cl}<0.065$) with dynamical state identification of `relaxed'($2$) or only those `mildly interacting'($3$) clusters which show a concentrated X-ray surface brightness distribution with mild optical and/or X-ray substructures but without obvious interaction features in the ICM (e.g ICM sloshing). Figure \ref{fig:sigz} summarises our cluster samples as a function of cluster velocity dispersion ($\sigma_{cl}$) and cluster redshift ($z_{cl}$) while Table \ref{tab:WINGS_clusters} enlists the global cluster properties of the samples. We describe each of the post-merger clusters as follows-

\begin{description}
    \item[\bf{A3667}] (Figure \ref{fig:tot_dist}; top left panel) is one of the most massive and complex merger systems in this sample with $M_{500}$=7.04$\pm$0.05$\times$10$^{14}$M$_\odot$ \footnote{\label{note2}SZ Masses obtained from Planck Union Catalog \citep{planck15}}. It shows evidence of recent major-merging activity through the presence of twin radio relics, with a bridge in between \citep{caretti13} and disturbed X-ray emission with cold fronts \citep{knopp96,mazzotta02,briel04,nakazawa09,fino10,sarazin16}. While the exact merger dynamics of A3667 are yet to be constrained, recent studies like \citet{sarazin16} propose a timeline of $\sim$1 Gyr since the off-axis pericentric passage of two equally massive systems occurring in the plane  of the sky. They also corroborate the merger shock being located at the outer edge of the stronger relic in the system ($\mathcal{M}\sim 2.5$), with a possible detection of another X-ray shock front ($\mathcal{M}\sim 1.8$) by the weaker relic \citep[][]{storm18}. The most detailed view of the radio relics in A3667 using MeerKAT Galaxy Cluster Legacy Survey \citep[MGCLS; ][]{knowles22} reveals a filamentary substructure in the stronger relic \citep{gasperin22}. Other follow-up investigations include characterisation of magnetic fields through the radio mini-halo  coinciding with one of the BCGs \citep{riseley15}, gamma-ray observations \citep{kiuchi09}, and optical weak-lensing studies \citep{joffre00}. 
    \item[\bf{A3376}] (Figure \ref{fig:tot_dist}; middle left panel) is one of the most well-studied extreme post-merger systems in the literature with $M_{500}$=2.4$\pm$0.2$\times$10$^{14}$M$_\odot$ \footnotemark[\value{footnote}]. It displays double relics, and asymmetric X-ray emission \citep[]{bagchi06,kale11,durret13,chibueze21,chibueze23}. The merger timescale for A3376 is $\sim0.6$ Gyr since pericentric passage, the shock front of which gave rise to a stronger ($\mathcal{M}\sim2.8$; $v_{\rm s}$=1630\,km\,s$^{-1}$) and a weaker relic \citep[$\mathcal{M}\sim1.5$, $v_{\rm s}$=1450\,km\,s$^{-1}$;][]{bagchi06,kale12,akamatsu12,lijo15,urdampilleta18}. Simulations \citep{machado13} constrain the plane-of-the-sky merger scenario with very low impact parameter ($\sim$ few kpcs), and a mass ratio of 3:1. Weak-lensing studies however reveal a more complex merging system with a third infalling group \citep{mo17}.
    \item[\bf{A168}] (Figure \ref{fig:tot_dist}; bottom left panel) is the lowest mass cluster ($M_{500}$=1.9$\pm$0.2$\times$10$^{14}$M$_\odot$ \footnotemark[\value{footnote}]) with a radio relic detection \citep{dwaraka18} and a radio halo with the least power discovered to date \citep{botteon21}. It shows an X-ray cold front \citep{hallman04} and two X-ray peaks detected in $Chandra$ data \citep{yang04}. They further speculate an off-axis merger of mass ratio of 1:1-3 and a shock speed of 600 km s$^{-1}$, putting the merger age to $\sim0.6$ Gyr.
\end{description}

In summary, although our post-merger cluster sample covers a wide range of cluster masses, it is homogeneous not only with respect to the dynamical states of clusters but also with respect to the merger timescales (TSC $\sim0.5-1$ Gyr). The homogeneity of the sample is extremely crucial to resolve conflicting inferences drawn from individual cluster merger studies from literature, and hence towards the first attempt at generalising galaxy post-processing in cluster mergers. 

\subsection{Galaxy sample}
\label{data-g}

The primary goal of this study is to investigate integrated star formation properties through galaxy colours, and linking them to the merger history of the clusters since $\sim1$ Gyr. Our base galaxy sample throughout this paper comprises all the galaxies with OmegaWINGS spectra. 

We use updated $B-$ and $V-$band magnitudes from the OmegaWINGS photometry recomputed with morphology-dependent $K-$corrections \citep{vulcani22}. Galaxy stellar mass is obtained as one of the data products of \textsc{sinopsis}\footnote{http://www.irya.unam.mx/gente/j.fritz/JFhp/SINOPSIS.html} \citep[SImulatiNg OPtical Spectra wIth Stellar population 
models;][]{fritz07,fritz17}, which reproduces the observed spectra using theoretical spectra of simple stellar population (SSP) models of 12 different ages - from $\sim10^{6}$ years up to the age of the Universe at the galaxy's redshift. We use stellar mass definition number 2 \citep[see][]{longhetti09} from the updated \textsc{sinopsis} outputs by \citet[][hereafter PM23]{pm23} obtained at cluster redshifts and using \citet{chabrier03} initial mass function (IMF), that includes stars in the nuclear-burning phase and remnants, but takes into account mass losses due to stellar evolution. This paper also presents one of the first comparative galaxy morphology demographics in cluster mergers, in extension to the qualitative results from \citetalias{kk20} for A3376. We use T-type galaxy morphologies computed by the tool \textsc{morphot} which uses a combination of 21 morphological diagnostics (derivable from imaging data), machine learning, and neural networks \citep{fasano12}. We broadly bin these T-type morphologies into 3 classes: ellipticals (E; --5.5 $< T_{M} <$ --4.25), lenticulars (S0; --4.25 $\leq T_{M} \leq$ 0), and spirals (Sp; 0$< T_{M} \leq$ 11) \citep{vulcani23}.

We also statistically account for the fact that not all galaxies detected
in the images have a spectroscopic counterpart, and have corrected for both radial (geometrical) $C(r)$ and
magnitude $C(m)$ completeness \citep{cava09,moretti17}. This is done by weighting the properties of each
galaxy with the product of the inverse of the two completeness values. Finally, we apply a stellar mass completeness limit of Log$(M_{*}/{\rm M}_{\odot})=9.48$ (corresponding to absolute $V$--band magnitude brighter than $M_{V}$ = --$18$ mag), and an additional radial cut of 0.9 $R_{200}$ for computing fractions to account for the different radial coverage for OmegaWINGS clusters (for more details see \citetalias{l22}). This radial cut roughly corresponds to the circumference of the post-merger shock front-- indicated by the radio relics-- presently observed for all the three cluster mergers. Hence, limiting the galaxy sample to this radius also ensures we are looking at the majority of the galaxies that are affected by the outgoing shock front, even though we would be sampling infalling galaxies (projection effects) and galaxies whose orbits have been modified by mergers.
We further reiterate that this sample presents updated galaxy morphologies and improved integrated magnitudes to those presented in \citetalias{kk20} for A3376, without affecting their results. Lastly, we use the \citet{wilson27} binomial confidence
interval to compute the $1\sigma$ uncertainty in the fractions presented throughout this paper.
\subsection{Galaxy distribution in post-merger cluster}
\label{data-gdist}

\begin{figure*}
    \centering
	
	\includegraphics[width=1.7\columnwidth]{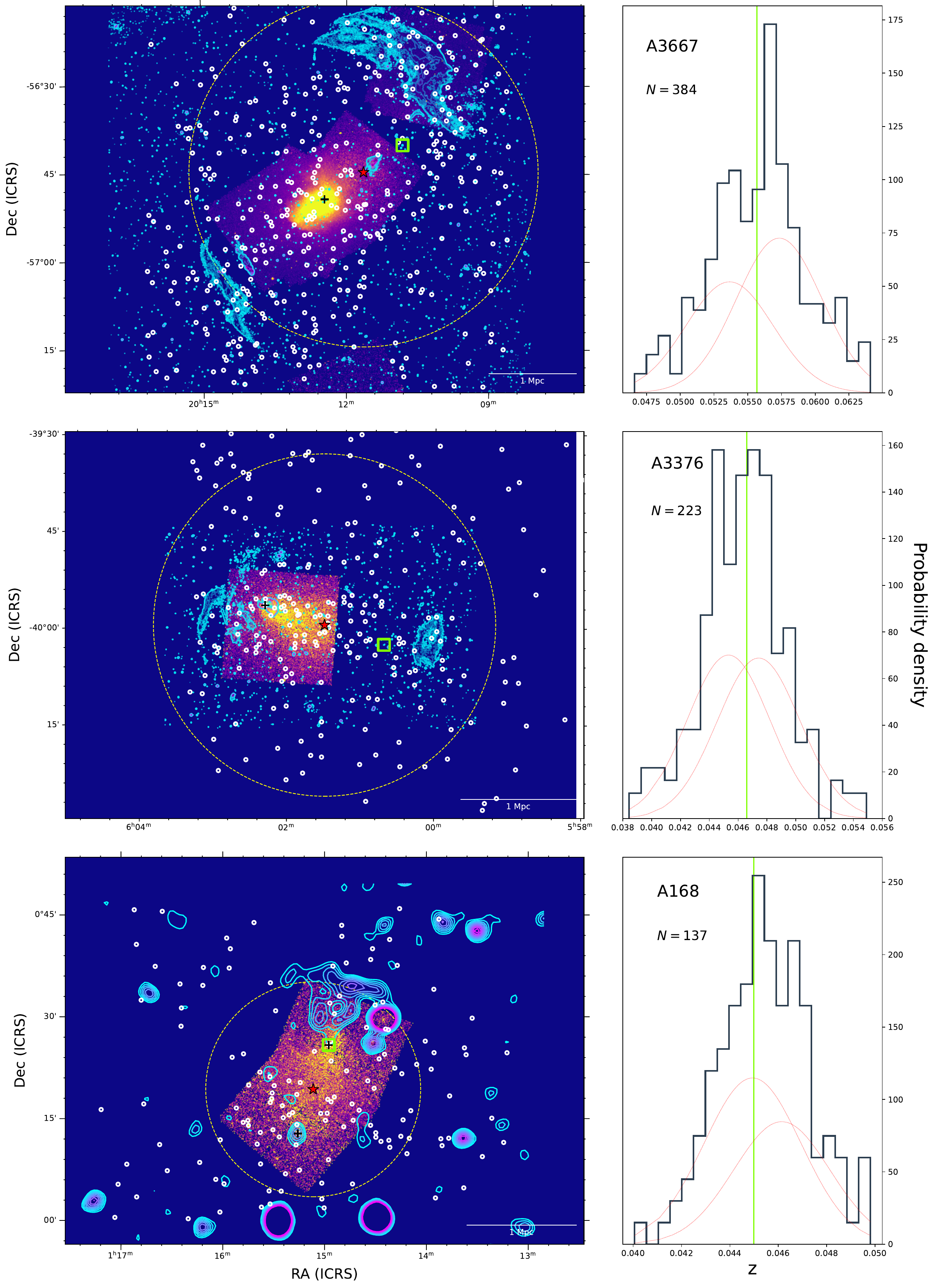}

    \caption{\textit{Left:} Spatial distribution of our spectroscopic galaxy sample (open white circles) of the three post-merger clusters in our cluster sample (labeled in the right-hand panels). BCGs are marked with open green squares and the centre of the system by a red star, both of which are adopted from \citetalias{l22}. The extended blue contours denote the radio relics using 1280 MHz data from the MeerKAT Galaxy Cluster Legacy Survey \citep[MGCLS,][]{knowles22} for A3667 \& A3376, and 170-230 MHz continuum data from the GaLactic and Extragalactic All-sky MWA survey \citep[GLEAM,][]{wayth15} for A168. The underlying image represents the cluster X-ray emission using the publicly available $Chandra$ X-ray data with black pluses indicating the peak X-ray emission. The overlay of the dotted circle indicates an aperture of 0.9 $R_{200}$. \textit{Right:} the spectroscopic $z-$distribution of the galaxy sample, and BCGs (green vertical line). The red peaks indicate the Gaussians fitted to the observed $z-$distribution while $N$ denotes the number of spectroscopic members in each cluster.}
    \label{fig:tot_dist}
\end{figure*}

A dynamically relaxed galaxy cluster displays a distinct distribution of galaxies where passive galaxies are most commonly found in dense cluster cores whereas star-forming galaxies prefer the relatively less-dense cluster outskirts \citep[]{gray04,gavazzi10,peng10,weinmann10,prescott11,wetzel12,haines13,davies16,grootes17,davies19,owers19}. This also gets translated into the observed morphology--density relation 
with galaxies showing early-type morphologies preferring dense cluster regions \citep[]{dressler1980,treu03,blanton05,bamford09,skibba09,george13,houghton15,fasano15,brough17,greene17,oh18}. A major interaction between galaxy clusters, however, is a disruptive event where the ICM, and the merging halos can result in a cluster-wide redistribution galaxies. Furthermore, the the presence of more than one radio relic indicates a plane-of-the-sky major merger suggesting an approximate mass ratio of at least $\sim$1:3 \citep{vweeren19}. 

We investigate the observed distribution of the galaxies as a function of both relic positions (and hence the shock-front) and the BCG \footnote{In this work, we define the BCG as the first brightest galaxy in the system as identified by \citetalias{l22}, which may not necessarily coincide with the central galaxies of the merging subclusters.} location in each of the three post-merger clusters A3667, A3376 and A168 (left panel in Figure \ref{fig:tot_dist}). We adopt the definition of the `centre' of the post-merger system from \citetalias{l22} and take the midpoint between the first BCG and the galaxy coincident with the X-ray peak. The choice of these galaxies takes into consideration the characteristic asymmetrical X-ray emission often observed in merging clusters, and most often associated with the core of the displaced subcluster. An in-depth discussion about the definition of cluster centres and BCGs for the three cluster mergers is presented in \citetalias{l22}. Spatially, A3667 and A3376 show fairly unimodal but elongated galaxy distributions aligning possibly along the axis of the merging event and of the two radio relics. This is corroborated by \citetalias{kk20} (and the references within) as the merger dynamics for A3376 are well constrained through simulations and weak-lensing analyses. A168 however shows galaxy distribution somewhat perpendicular to the direction of the BCGs and the relic.

Further insights are given by the $z-$distribution (right panel in Figure \ref{fig:tot_dist}) which shows distinct peaks for A3667 indicating a close encounter with non-negligible impact parameter, the broad peak for A3376 and a redshift distribution mimicking a relaxed cluster for A168. Except for A3667, the $z-$distribution for each of our clusters can be fitted by double Gaussian peaks but with minimal separation. This qualitatively corroborates a close encounter between galaxy clusters leading to an unimodal distribution of galaxies which was perhaps more disruptive to the galaxy distribution in A168 as compared to that in the other cluster mergers. This can be expected as A168 relatively has a lower mass for a major merger. Furthermore, we also corroborate a headlong collision between two massive halos for A3376, confirmed through simulations. We performed a Shapiro-Wilk normality test \citep{shapiro65} under the null hypothesis that the observed $z-$distribution for each of the clusters is drawn from a Gaussian distribution. Out of the three cluster mergers, only A3667 yields a Shapiro-Wilk $p-$value of 0.057 indicating a possible non-Gaussian distribution of galaxies. We also performed a Hartigan's dip test \citep{hartigan85} under the null hypothesis that the observed $z-$distribution for each of the clusters is unimodal. With neither of the clusters yielding a $p-$value $<$0.05, we can conclude that all the three post-mergers display unimodal $z-$distribution. Thus, even if we are sampling merging events with similar TSC, it is evident that the masses of the merging halos together with the dynamics play an important role in the final configuration observed of the whole merging system. 

\section{Galaxy morphology fractions as a function of post-merger cluster environment}
\label{sec:morph}

\begin{figure*}

	\includegraphics[width=18.5cm]{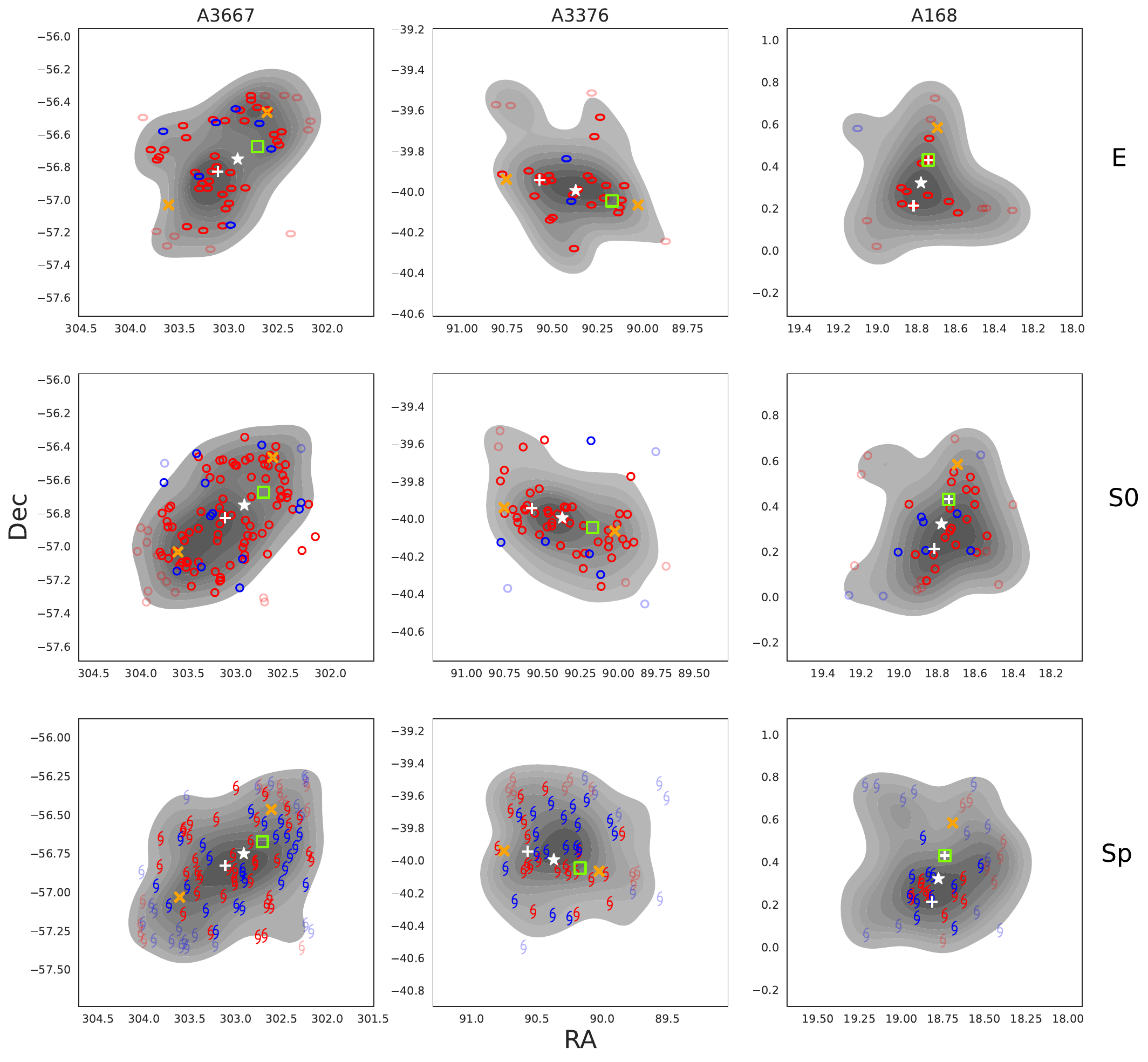}
    \caption{Spatial distribution of galaxies for each of the morphologies - ellipticals (top), lenticulars (middle) and spirals (bottom) from the primary spectroscopic sample. Each panel shows the overall density contours for the distribution of each morphology type, split into red and blue galaxies according to the selection criteria discussed in Section \ref{sec:gcolors}. The faint markers indicate galaxies beyond 0.9$R_{200}$ and are thus excluded from the quantitative analysis presented in this paper. As in Figure \ref{fig:tot_dist}, we show the BCG (open green square), system centre (white star), X-ray peak (white plus) and approximate location of radio relics (orange crosses).      }
    \label{fig:morph_dist}
\end{figure*}

We split the different morphology classes as defined in Section \ref{data-g} and inspect their spatial distribution in Figure \ref{fig:morph_dist}. The over-densities of ellipticals of the merging subclusters appear to be relatively undisturbed in A3667. The picture is somewhat extreme and opposite for A168 where elliptical galaxies -- however low in number-- appear to be situated along the line joining the relic with BCGs and also perpendicular to it. This trend is mirrored in S0s of A168 while A3667 shows an elongated/narrow S0 distribution without individual over-densities. Spiral galaxies on the other hand display a very narrow/flattened distribution (akin to S0s) in A3667 while clusters A3376 \& A168 show wider spatial distribution for spirals. However, it should be noted that the spatial distribution of spiral-type galaxies would be much affected by projection effects and the fact that we will be sampling extra cosmic web.

\begin{figure}
    
	\includegraphics[width=1\columnwidth]{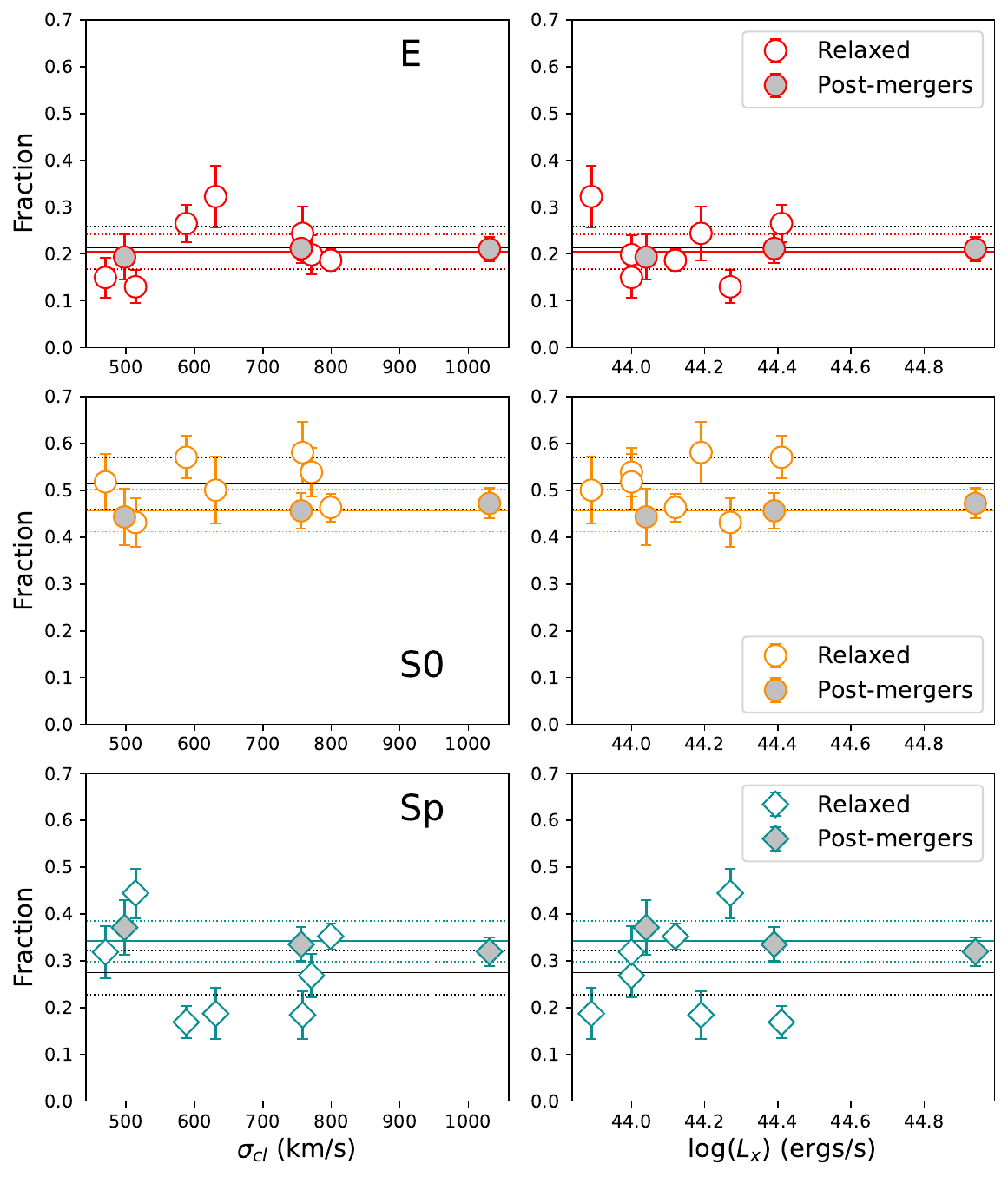}

    \caption{Fraction of galaxies of each morphology as a function of $\sigma_{cl}$ (left) and X-ray luminosity ($L_{x}$) (right). Filled grey symbols indicate the clusters in the post-merger cluster sample. The solid coloured and black lines denote the average fractions in the post-merger, and control cluster samples respectively, with errors represented by the corresponding dashed lines. Cluster mergers demonstrate a constant fraction of spirals in comparison to the seven relaxed clusters. }
    \label{fig:frac_morph}
\end{figure}

Morphology fractions give us important clues not only to the build-up of cluster galaxy populations over time ($z$-dependence) but also plausible cluster environment-specific processes which could lead to morphological transformations of galaxies \citep[dependent on global cluster properties; e.g.][]{desai07,poggianti09,vulcani11a,vulcani11b,calvi12}. We revisit the morphology fractions under post-processing due to cluster mergers (Figure \ref{fig:frac_morph}) and examine fractions of E, S0 and Spiral galaxies respectively as a function of $\sigma_{cl}$ (and hence the cluster-mass range of our sample) and cluster X-ray luminosity ($L_x$). We further separate our post-merger cluster sample (grey filled markers) from the relaxed cluster control sample (open symbols). We quantify the comparison using average fractions for the entire post-merger (denoted as coloured solid lines) and the control cluster sample (denoted as black solid lines) with 1$\sigma$ error limits (dotted lines of respective colour) for each morphology type. These plots recreate the results from \citet{poggianti09} who use the entire WINGS sample but with a cut in absolute $V-$band mag $<-19.5$) and aperture (0.6 $R_{200}$), and \citetalias{pm23} who extend it to include the entire OmegaWINGS cluster sample with the same limits. We reiterate that we use a mass-complete spectroscopic sample within 0.9 $R_{200}$ fixed aperture to compute the morphology fractions. 

This work is one of the first studies exploring morphological fractions in clusters that are characterised by profoundly different dynamical states and growth history. Even though our primary cluster sample is comprised of only three post-merging systems, we demonstrate that the fractions of Es and S0s remain unchanged irrespective of the dynamic stage of our sample or the range of $\sigma_{cl}$ or $L_x$. Consequently, we report a constant fraction of spirals in post-mergers (0.34$\pm$0.04) similar to relaxed clusters (0.27$\pm$0.04). However, the spiral fraction in relaxed shows considerable scatter both with respect to $\sigma_{cl}$ \&  $L_x$, making their fraction inconclusive for direct comparison with that of post-merger clusters. The constant spiral fraction for post-merger clusters demonstrates a fairly mixed population of spiral galaxies, now accreted in the post-merger systems either as a part of original merging subclusters or the extended cosmic web. A detailed comparison as a function of cluster-centric radius or projected phase--space however is not possible as the system centres of merging clusters are not exactly equivalent to the centres of relaxed clusters. Nonetheless, this poses newer challenges to our understanding of the morphological evolution of cluster galaxies due to probable post-processing because of extreme dynamical galaxy environment.  
 
\section{Galaxy colours: Integrated star formation properties of galaxies in post-merger clusters}
\label{sec:gcolors}
We use galaxy colours as a measure of integrated star formation and divide our spectroscopic sample into red and blue galaxies. This will provide a general idea about the recent star formation in post-merger clusters, and their comparison with the expected trends in relaxed clusters will confirm whether cluster-merging activity influences the star formation properties of galaxies. 
\subsection{Colour--magnitude relation of merging clusters}
\label{subsec:cmr}

\begin{figure}      
    
    \includegraphics[width=1.1\columnwidth]{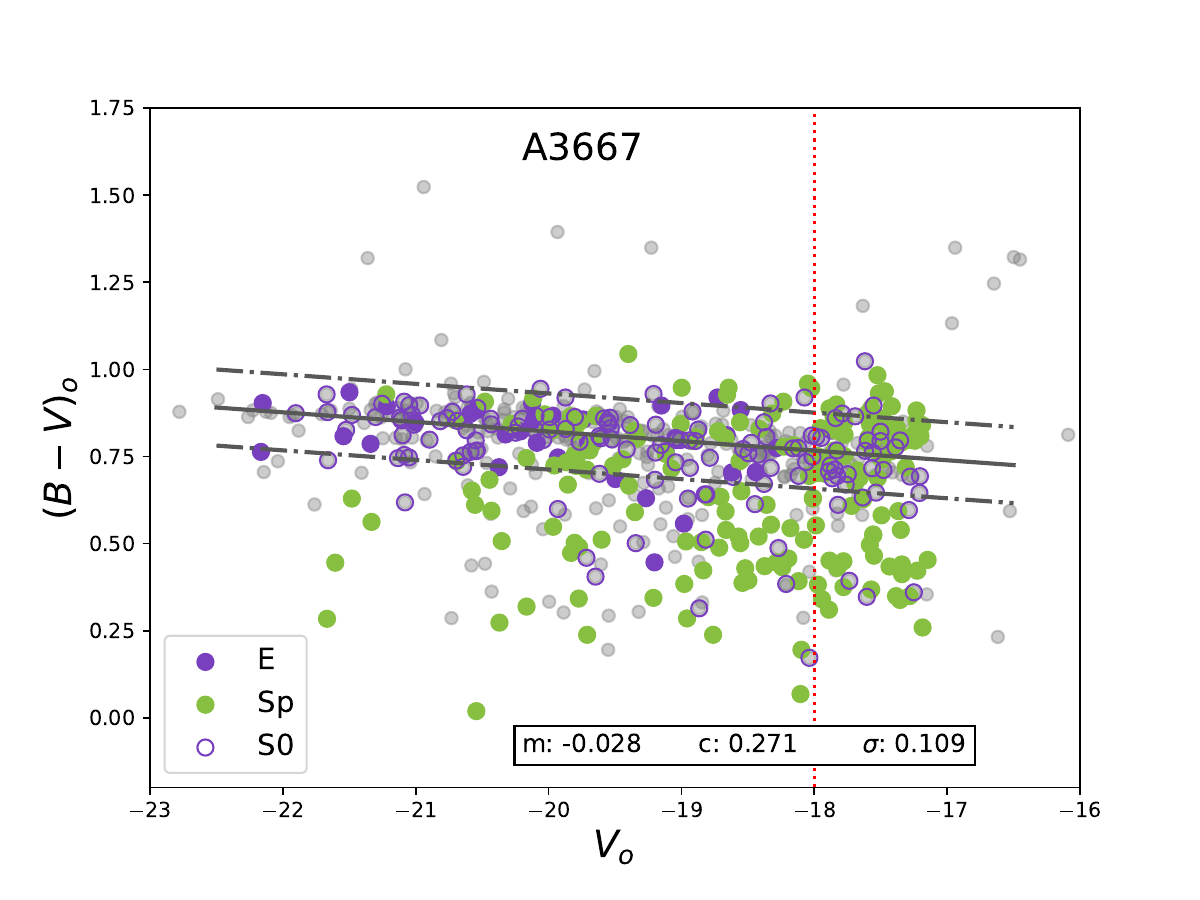}%
    \vspace{-2\baselineskip}
    \includegraphics[width=1.1\columnwidth]{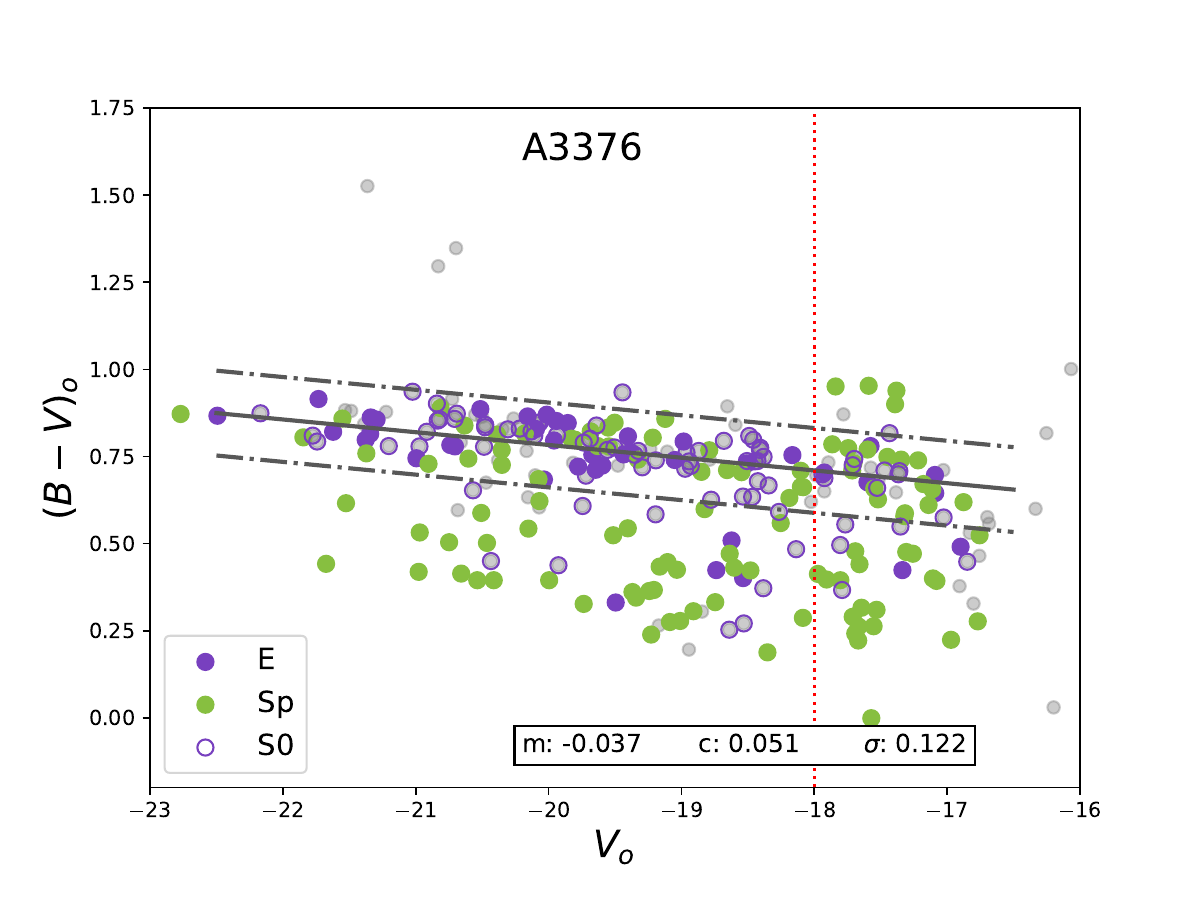}%
    \vspace{-2\baselineskip}
    \includegraphics[width=1.1\columnwidth]{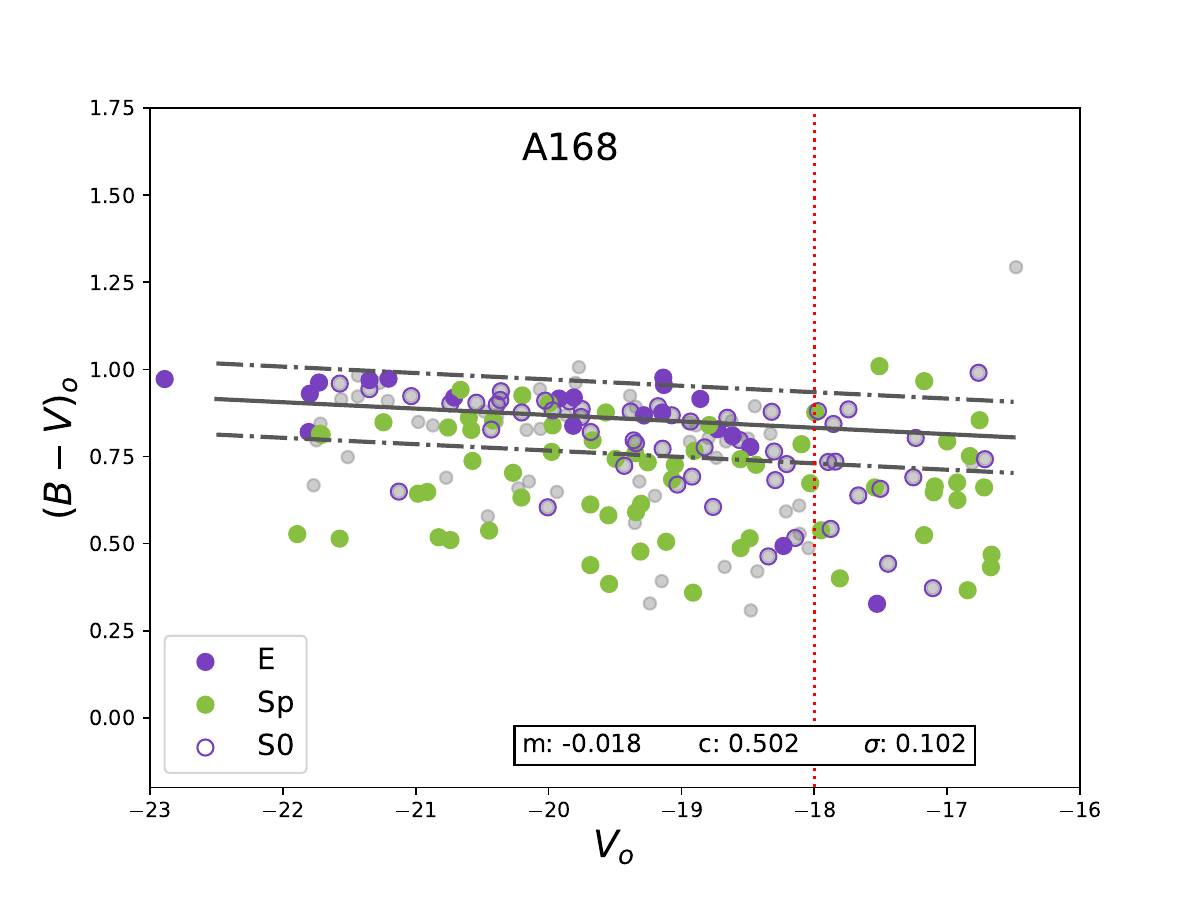}
    \caption{Colour--magnitude relation for A3667, A3376 \& A168 using the OmegaWINGS photometric sample (grey filled circles). Overplotted are galaxies in the extended spectroscopic galaxy sample \citep{vulcani22} with \textsc{morphot} T-type morphologies. The dash-dotted line represents the 1-$\sigma$ above and below the fitted red sequence (solid) while the vertical dotted red line indicates the $V-$band magnitude limit ($V_0=-18$) corresponding to the spectroscopic mass-completeness of log$(M_{*}/{\rm M}_{\odot})=9.48$. The inset box indicates the slope (m), intercept (c) and the standard deviation ($\sigma$) of the linear fit to the red sequence.} 
    \label{fig:cmr}
\end{figure}

\begin{figure*}
    \centering
    
    \hspace{-0.9cm}
    \includegraphics[width=0.7\linewidth]{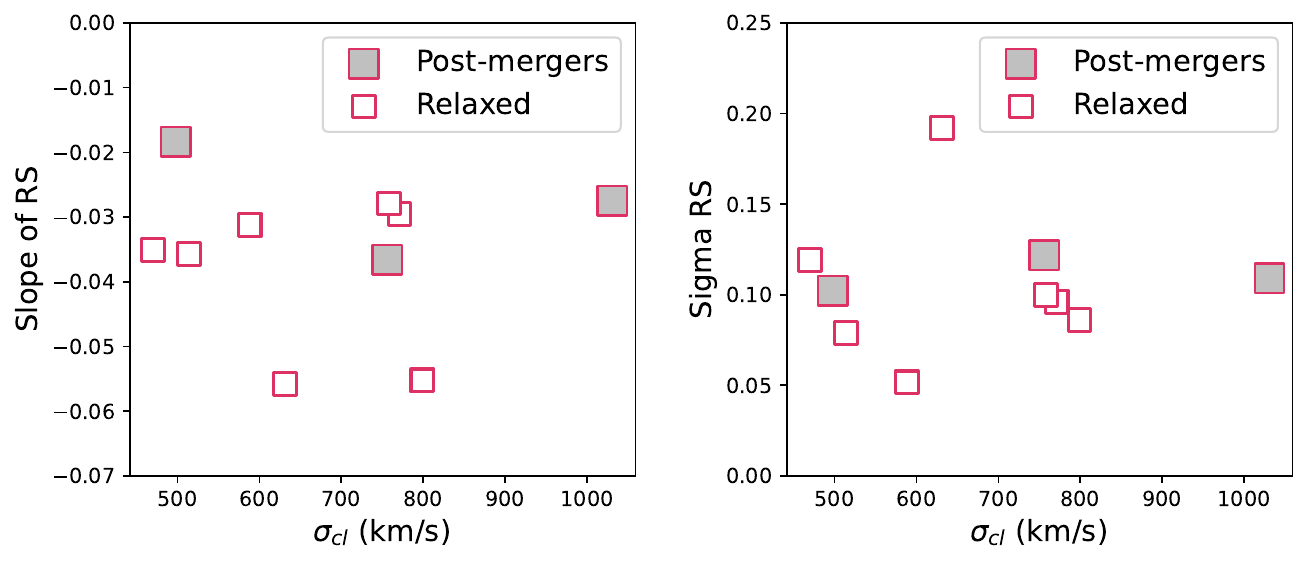} 
    \caption{Slope of RS ($m$) as a function of $\sigma_{cl}$ (left) and scatter of RS ($\sigma$) as a function of $\sigma_{cl}$ (right). Filled grey symbols indicate the post-merger cluster sample. Both relaxed and cluster mergers show similar slope and scatter.} 
\label{fig:rs_prop}
    

\end{figure*}

\begin{figure*}
     \centering
     \begin{subfigure}[b]{0.35\textwidth}
         \centering
         \includegraphics[width=\textwidth]{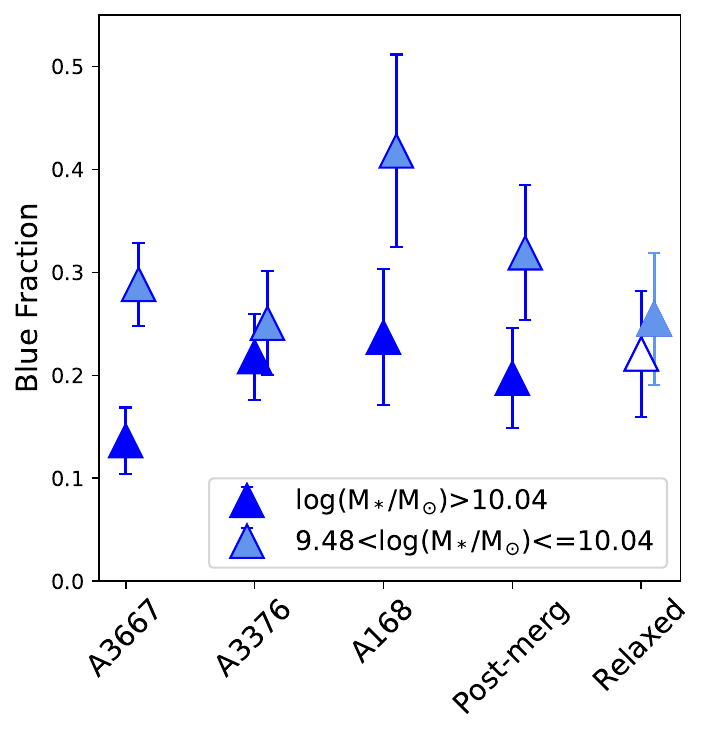}
     \end{subfigure}
     \hspace{0.5cm}
     \begin{subfigure}[b]{1.1\columnwidth}
         \centering
         \includegraphics[width=1.1\columnwidth]{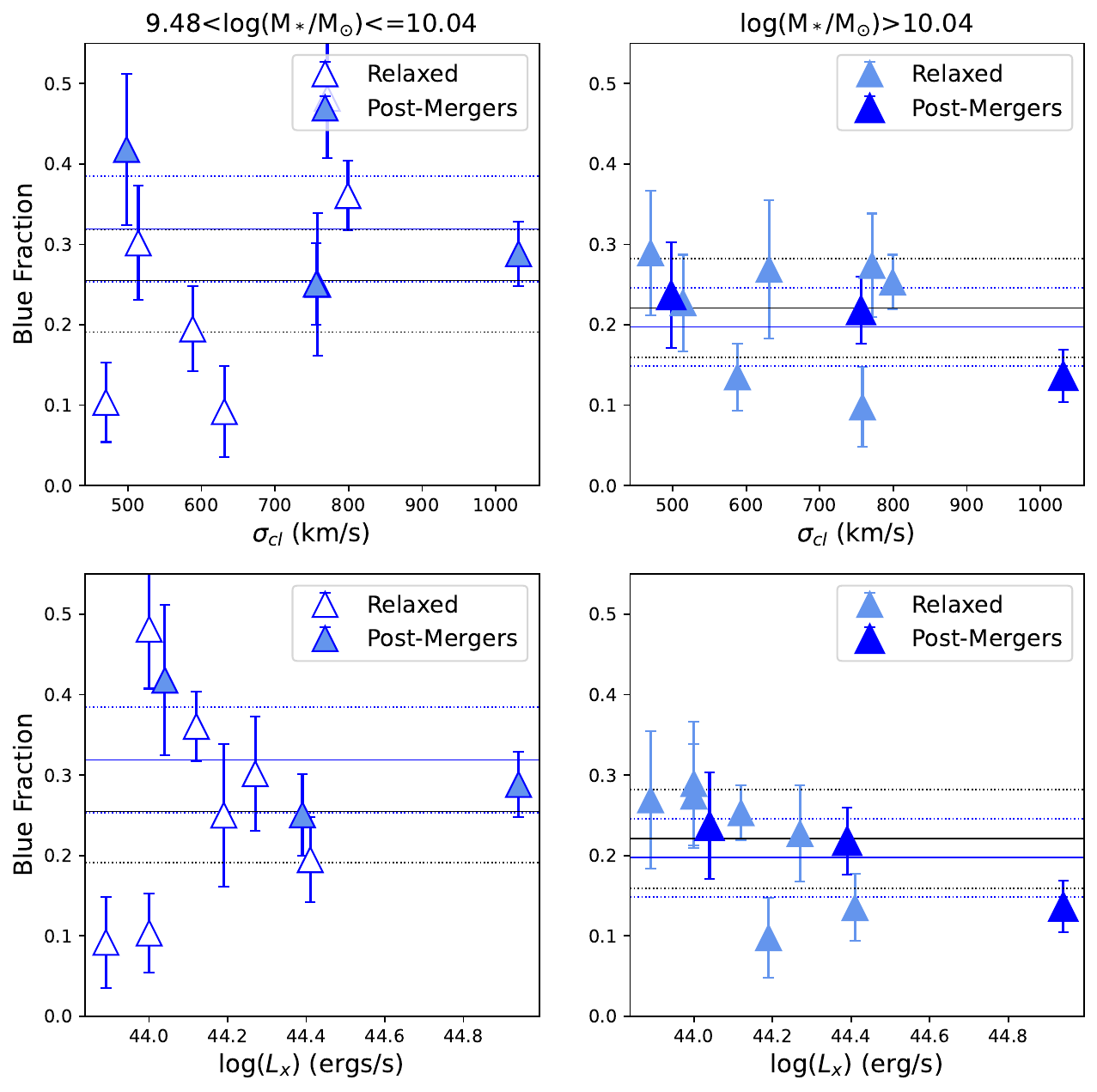}
         
     \end{subfigure}
     \hspace{0.9cm}
     
     \caption{ Blue fraction as a function of cluster dynamical state, in two mass bins. \emph{Left}: average blue fraction as a function of cluster dynamical state, in two mass bins. \emph{Right}: Blue fraction as a function of $\sigma_{cl}$ (top) and $L_{x}$ (bottom) for each of the two mass bins. The solid and dotted black and blue lines represent the average fractions with 1-$\sigma$ errors for relaxed and cluster mergers respectively. We observe the blue fractions, dominated mostly by spiral galaxies, to be similar in post-mergers and relaxed clusters for the lower mass bin. Moreover, the blue fractions as a function of global cluster properties remain constant within 1$\sigma$ errors for post-mergers. The relaxed clusters on the other hand show a lot of scatter in the blue fractions as a function of $\sigma_{cl}$ \& L$_{x}$}
     \label{fig:frac_blue}
\end{figure*}

To identify red and blue galaxies, we use the extended OmegaWINGS spectroscopic sample, compiled by \citet{vulcani22} which includes the extra spectroscopic redshifts they compiled from the literature. We (i) construct a rest-frame $(B-V)_{0}$ colour--magnitude relation (CMR) individually for each cluster merger, (ii) fit the red sequence (RS) for the entire sample irrespective of galaxy morphology, (iii) use a boundary of 1-$\sigma$ below the RS to select red (those lying above this boundary) and blue (below this boundary) galaxies. We repeat the same procedure for the control cluster sample and use individual RS-fits to take into account cluster-to-cluster variation. 

For the RS-fitting, we fit a double-Gaussian to the distribution of $(B-V)_{0}$ colour in four $V_0$ magnitude bins (\citet{franco23}; See also \citet{jake17}). The RS is then defined by a linear fit across the  four peaks of the redder Gaussians in each of the four magnitude bins, and the average scatter of each cluster is taken as the standard deviation $\sigma$. Figure \ref{fig:cmr} show the CMRs for A3667, A3376 and A168 respectively along with the RS fit parameters of slope ($m$), intercept ($c$) \& standard deviation ($\sigma$). The CMRs for control clusters are presented in Appendix \ref{ap:cmr-control}. Note that these,  while being an improvement with updated magnitudes and K-correction, are directly comparable to those presented in \citet{valentinuzzi11} but over OmegaWINGS field of view.  

From Figure \ref{fig:cmr}, it is clear that A3667 already has a densely populated red sequence contributed by the massive merging subclusters. For A3376, we present a direct improvement to the CMR presented in \citetalias{kk20} due to updated magnitudes and hence updated RS-fitting. A168 being the lowest-mass cluster merger clearly shows a red-sequence `in-progress' especially at magnitudes brighter than $V_0=-18.5$. We further compare  $m$ and $\sigma$ between post-merger and relaxed cluster samples (Figure \ref{fig:rs_prop}). We find that all three post-merger clusters show the least variation in their $m$ and $\sigma$. For the control relaxed cluster sample, the minimal scatter in $\sigma$ could highlight the different formation histories of the clusters, with some of them being young and still building their red sequence. 

\subsection{ Blue fraction in merging clusters }
\label{ssubsec:blue_frac}

\begin{figure*}
	\hspace*{1.8cm}
	\includegraphics[width=0.8\textwidth]{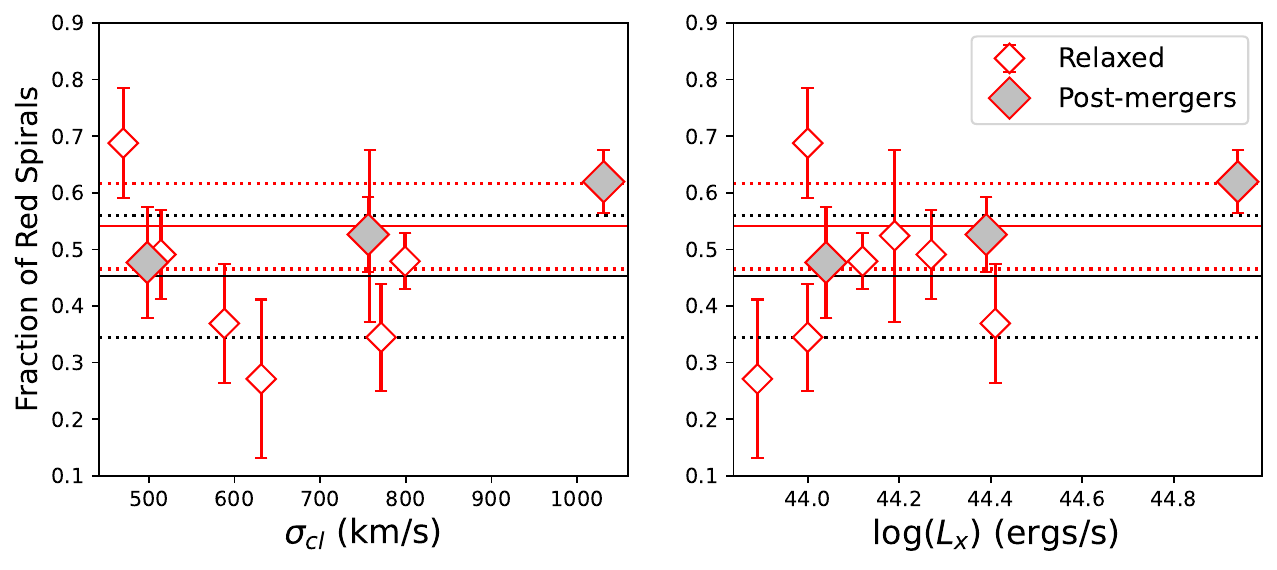}
    \caption{Fraction of red Spirals as a function of $\sigma_{cl}$ and $L_{x}$, separated into relaxed cluster sample (red open diamonds) and merging cluster sample (filled grey diamonds). The solid black and red lines indicate the average fraction for relaxed and merging clusters respectively. The dotted lines correspond to the 1$\sigma$ error for each respective sample. A3667 demonstrates a marginally higher fraction of red spirals as compared to relaxed clusters. }
    \label{fig:frac_redsp}
\end{figure*}

The existence of a galaxy population in transition has long been considered proof of delayed environmental effects affecting galaxy structure and star formation on different timescales. Indeed, studies have found that galaxy clusters host non-negligible populations of red passive spirals thereby supporting cluster-specific physical mechanisms like ram-pressure stripping to be the cause of quenching of star formation while leaving galaxy structure intact \citep[e.g.][]{bamford09,kk17,bremer18}. Upon a major cluster merger however, galaxies would further undergo environmental `post-processing' as a result of such energetic dynamic activity which could lead to further changes in the star formation of galaxies already affected by the dense cluster environment before cluster-cluster merging.

We first start by exploring blue fractions in post-merger clusters, which have never been explored before directly in comparison to the relaxed cluster environment. Figure \ref{fig:frac_blue} compares the fraction of blue galaxies (all morphologies) in two stellar mass bins between the three post-mergers and the seven relaxed clusters, computed within a fixed aperture of 0.9$R_{200}$. The mass bins are selected based on the median stellar mass of blue galaxies of our entire post-merger+relaxed cluster sample -- dominated by spiral galaxies -- because they show similar stellar mass distribution irrespective of the cluster dynamical stage (Figure \ref{fig:a2}).

Figure \ref{fig:frac_blue} (left panel) shows that the average blue fractions in post-mergers are comparable to those in relaxed clusters throughout the stellar mass range. However, the average blue fraction for post-mergers appears to be dominated by the excess of blue galaxies in the lower stellar mass bin for A168 and the lack of high-mass blue galaxies in A3667. The blue fractions for relaxed clusters as a function of global cluster properties like $\sigma_{cl}$ \& $L_{x}$ (Figure \ref{fig:frac_blue}b) show significant scatter, especially in the lower mass bin. This prevents us from confirming whether the elevated blue fraction in A168 is a post-processing signal. This also calls attention to the cluster-to-cluster variations within the cluster merger sample and highlights the biases that may be introduced in single-cluster case studies. For example, the analysis of A3376 by \citetalias{kk20} concludes that A3376 appears to have star formation comparable to relaxed clusters which can also be indirectly concluded from Figure \ref{fig:frac_blue}a. Taken together with the constant fraction of spirals in post-merger clusters (Figure \ref{fig:frac_morph}), the average comparable blue fractions for post-merger \& relaxed clusters could suggest a levelling of galaxy colours contributed through the newly accreted galaxies, now part of the post-merger system \citep[See also][]{mulroy17}. 

We next inspect the spatial distribution of the galaxies comprising the blue fraction and the red galaxies separated by broad morphologies, for all three cluster mergers. Figure \ref{fig:morph_dist} highlights the galaxies within 0.9$R_{200}$, to enable a direct association with the blue fractions. We note that the majority of the red early-type galaxies (E or S0) in all three cluster mergers follow the  general direction along the line joining the approximate relic positions with the BCGs, with an exception of red ellipticals in A168 (Refer to Section \ref{sec:morph}). 
Spirals, irrespective of their colours, appear to be widely distributed. We thus conclude that red and blue galaxies in post-merger clusters do not demonstrate any spatial preference. However, we reiterate that obtaining accurate phase--space information for merging clusters is difficult due to extra projection effects and merger configuration.    

\subsection{Red spirals in cluster Mergers}
\label{ssubsec:red_sp}

\begin{figure*}
	\centering
	\includegraphics[width=0.8\textwidth]{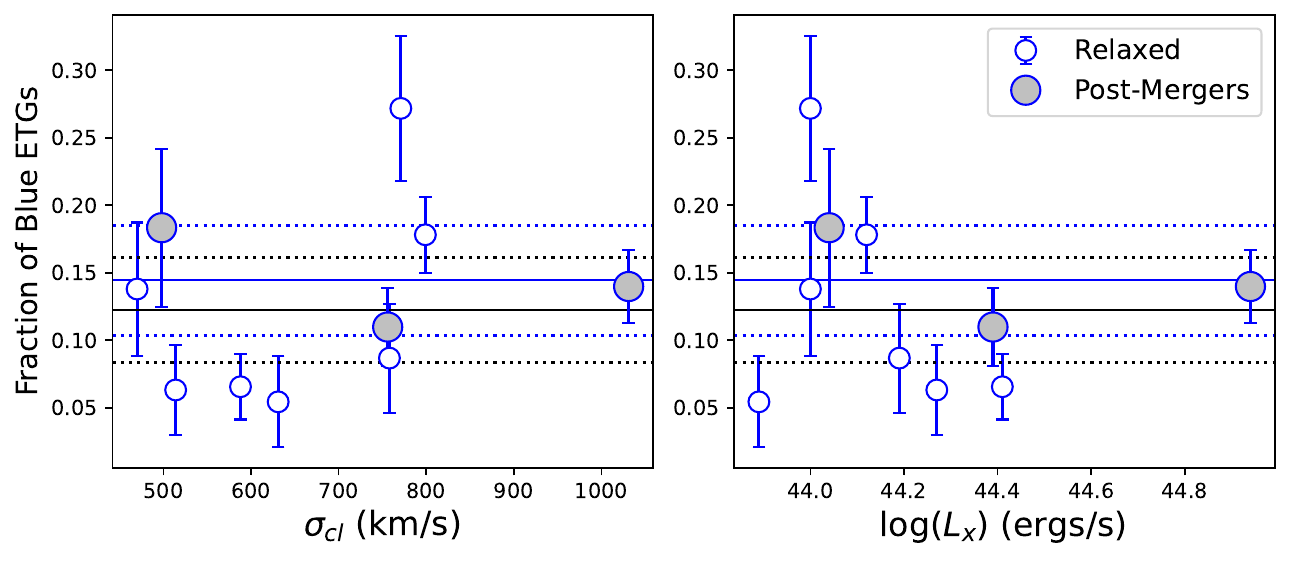}
    \caption{Fraction of blue ETGs as a function of $\sigma_{cl}$ and $L_{x}$, separated into relaxed cluster sample (blue open circles) and merging cluster sample (filled grey circles). The solid black and blue lines indicate the average fraction for relaxed and merging clusters respectively. The dotted lines correspond to the 1$\sigma$ error for each respective sample. Merging clusters show a constant fraction of blue ETGs while the relaxed clusters demonstrate considerable variation in the fraction of blue ETGs. }
    \label{fig:blue_lent}
\end{figure*}

Another signature of the pre-merger cumulative environment in post-merger cluster systems is the incidence of red or passive spirals in clusters, which \citetalias{kk20} proposed qualitatively using A3376 in their pilot study. We test this quantitatively by looking at the fraction of red spirals in all three cluster mergers i.e. the number of red spirals / total number of spirals, and comparing them to that in the relaxed clusters (Figure \ref{fig:frac_redsp}). The red spirals were selected as those above the RS-1$\sigma$ boundary, individually for each and every cluster. We discover that on average post-mergers show a comparable fraction of red spirals compared to that in relaxed clusters, except for A3667 which shows a marginally higher fraction than the average fraction for the relaxed clusters. However, this average fraction of $\sim45$\% for relaxed clusters is at least twice the fraction observed by \citet{valentinuzzi11} of $\sim20$\%, who use the entire WINGS cluster sample but with photometric galaxy memberships and smaller spatial coverage.

Edge-on spirals can also display red colours due to dust attenuation which we have not controlled for in either of our samples. However, a basic visual check using the publicly available Legacy Survey DR10 \citep{dey19} images renders a comparable fraction of nearly edge-on spirals in both the samples ($\sim10\%$). We do note a higher incidence of red spirals with a smooth disk \citep[no prominent spiral arms; See also][]{kk17} especially in A3667, compared to a general mixed collection of late-type spirals with strong bulges/smooth disk galaxies with possible S0 morphologies/barred galaxies across both post-mergers and relaxed clusters' red spiral sub-samples. This highlights that S0 vs. spiral classification is not a clean dichotomy and contamination of either morphological sample can occur especially when considering such galaxy population in transition. We further confirm a mean inclination of $\sim$ 48\degree\hspace{0.05cm}  for the red spirals in both our cluster samples (Figure \ref{fig:a3}), computed using the axis ratios from the OmegaWINGS $B-$, $V-$ photometry catalog \citep{gull15}.

Optically passive or red spirals have been encountered in clusters at both low-$z$ \citep[e.g. ][]{wolf09,valentinuzzi11,vulcani15} and intermediate-$z$ \citep[][]{sblazquez09,kk17}. Studies such as \citet{bamford09} show a higher variation in the fraction of cluster red galaxies than in the fraction of cluster early-type galaxies, proposing galaxies retaining their spiral morphologies while turning red/passive. All these findings support a rapid gas removal physical process (e.g. ram-pressure stripping, thermal evaporation), for the quenching of star formation with a delayed morphological transformation \citep[See also ][]{lopes16,mahajan20}. While optical colours could indicate their passivity, studies such as \citet{wolf09} found these optically passive red spirals at $z\sim0.17$ to have substantial star formation albeit lower than cluster blue spirals, and predominantly incident within $10<$log$M_*/M_{\odot}<11$. For cluster mergers however, this would form the pre-merger cumulative environmental influence as a result of galaxies already experiencing high-density environment before the cluster merging activity and hence would be a major contributor in the observed trends in the red spirals in our merging cluster sample. On the other hand, the hydrodynamic changes in the environment post cluster merger could lead to accelerated quenching, rendering the member galaxies passive.  

However, a deeper understanding of the stellar ages of these odd red spirals in both post-mergers and relaxed clusters is essential in order to confirm the origin of such populations ( effect of cumulative cluster environment or merger-driven) and the plausible physical processes involved with the timescales over which these are observable (Kelkar et al.; in prep).

\subsection{Blue ETGs in merging clusters}
\label{ssubsec:blue_lent}

Believed to be a class of evolved spirals, recent studies have confirmed that lenticular galaxies in both cluster and field environments result from multiple formation pathways \citep{ej21}. Structurally, they are characterised by a prominent bulge with a fading disk and older stellar populations, though they are found to be a lot more complex than this \citep[e.g.][]{ej22b}. This makes them a key population in understanding the transformation of spiral galaxies in high-density environments \citep[See also][]{jaffe11b}. Extending to general early-type galaxy population, studies such as \citet{bamford09} show that blue ETGs are low-mass galaxies preferentially located in low-density environments. We explore fractions of blue early-type galaxies, encompassing S0s and Es in our sample, in both merging and relaxed cluster environments.  

We discover that post-merger clusters seem to host a minor but significant population of blue ETGs (0.14$\pm$0.04). Such a blue ETG population is minimal for the majority of the relaxed clusters in our sample (Figure \ref{fig:blue_lent}), with an exception for A3560 (0.18$\pm$0.03) and A151 (0.27$\pm$0.05). These observations lend newer insight into the origins of such exotic populations in merging clusters. The bluer colours of early-type morphologies like S0s could simply indicate rejuvenated star formation due to enhanced galaxy interactions, which the large-scale merging activity could enable. For example, \citet{ej14} find bulges of S0s in Virgo and Fornax clusters host younger stellar population \citep[see also ][]{jaffe14}.
On the other hand, the general blue ETG population may be introduced in the merging cluster systems through the extended cosmic web (e.g. filaments). This is supported by the higher fraction of blue ETG galaxies for A3560 which, despite showing characteristics of relaxed clusters, is part of the broader Shapley Supercluster system and hence connected with the cosmic-web network. Furthermore, findings such as \citet{dhiwar23} find that blue ETGs with MilkyWay-like stellar masses (log$M_*/M_{\odot}\sim10$) reside in low-density environments, making them a probable population of cosmic filament-like environment now observed as a part of the newly forming post-merger system. Despite lacking specific stellar population information from these galaxies, the forthcoming follow-up work (Kelkar et al., in prep) explores this through luminosity-weighted ages in the central regions of these galaxies. 

\section{Discussion}
\label{sec:disc}

This comparative analysis is geared towards bridging the gap between individual cluster studies looking at effects of merger hydrodynamics (e.g. merger shock) of different timescales on star formation properties, and cluster population studies which encapsulate varied cluster dynamical ages observable over a few Gyrs to collectively infer the effect on galaxy evolution in extremely violent environments. 
Our explorations into the morphology fractions and galaxy colours within three young post-merger systems reveal a complex effect of the cumulative environment prior to the major merger. 

Dynamically young merging cluster systems are expected to be located at extreme cosmic web nodes, engulfed by dense large-scale environment of cosmic filaments feeding into the merging system. By construction, therefore, these systems are expected to be highly evolved with respect to their galaxy populations. We find evidence of this through the analysis of CMRs where post-merger clusters in our sample demonstrate constant scatter suggesting that red-sequence galaxies were already in place prior to such large-scale interactions. The variation in the blue fraction of relaxed clusters further proves that the comparison control sample is in fact an assortment of clusters with different global star-formation histories. On the other hand, simulations of cluster mergers have already demonstrated that the notch-like features in the outer edge of post-merger shock -- observed in both A3667 and A3376-- are likely indications of the shock-front interacting with the attached filament network surrounding the merging cluster system \citep{paul11}.
The network of cosmic web surrounding merging cluster systems would also result in a significant influx of new galaxies being introduced in the post-mergers in addition to the global redistribution of galaxies taking place throughout the merging event. The constant fraction spiral galaxies encountered in merging clusters, the levelling of blue fractions, and the incidence of exotic populations like blue ETGs could be a result of this. Our study provides an additional highlight that post-merger cluster systems with radio relics can thus be unique systems to explore such galaxy populations in filaments.

Galaxy transformations working on longer timescales further add an interesting arc to the story of the post-processing of galaxies in post-merger clusters. The general incidence of red spirals in both relaxed and post-merger clusters is likely a result of the cumulative pre-merger environment. However, the variation in galaxy properties presented by dynamically relaxed clusters, limits our ability to identify and confirm any possible post-processing signal these cluster mergers demonstrate such as the elevated and suppressed blue fractions in A168 and A3667 respectively, and the elevated red spiral fraction in A3667. Studies such as \citet{im02} demonstrated that galaxies with smooth morphologies can take $\geq$1 Gyr to turn red following a starburst. Physically, the $\sim$1 Gyr timescale could therefore be too short to identify the immediate effects of merger hydrodynamics on the colour and structural transformation of galaxies, as opposed to direct star formation measured in galaxies.  Either way, this study underlines the crucial importance of accounting for the cumulative environmental influences galaxies would likely undergo when hunting for galaxies directly affected by the major cluster merger. This can potentially bias previous observations reporting enhanced star formation in cluster mergers with radio relics \citep[e.g.][]{ferrari06,sobral15,stroe17,stroe21} while lending support to any `excess' in star formation rates being attributed to freshly accreted galaxies \citep[e.g.][]{chung10}. 

The uniformly matched coverage of 0.9$R_{200}$, while being smaller for the typical physical scale of post-merger clusters, gives our study two unique benefits - (i) it targets the central merging body of the system but also approximately covers the area within shock-fronts ensuring their connection to the plausible post-processing signatures and (ii) it allows for a statistical comparison of both the cluster samples across the same physical region. This approach differs from similar studies so far, specifically \citet{stroe21} which explores the wider infalling environment and \citet{mansheim17-b} who utilised the super-field as a comparison dataset for the galaxies in the Musket Ball cluster. 

Lastly, our analysis is not devoid of caveats which will influence our results to a certain degree, namely - (i) projection effects will always play a role even if the configuration of post-merger clusters with radio relics will almost always be in the plane of the sky (ii) these systems have not dynamically coalesced into a single final system making it difficult to take advantage of quantitative cluster galaxy distributions like the projected phase--space and local density measures. Redefining `centres' of cluster mergers has been a resolving step in this direction, which no literature study involving post-merger clusters with radio relics and their effects on galaxies' star formation, has attempted. Time-sensitive stellar diagnostics would thus be the key to further targeting any plausible signal of post-processing and consequently identifying specific merger-related hydrodynamical processes involved, which will be explored in the follow-up work (Kelkar et al., in prep). 

\section{Conclusions}
\label{sec:conc}

We present a detailed analysis of galaxy morphologies \& colours in three nearby ($0.04<z<0.07$) young ($\sim$0.6-1 Gyr) post-merger cluster systems --A3667, A3376 \& A168-- and seven complementary relaxed clusters. Exploiting the spectroscopic and  photometric data from OmegaWINGS, our investigations were able to present the complex nature of the cumulative environmental effects leading up to the merging event through exotic cluster populations such as red late-type and blue early-type galaxies. The primary conclusions so far are :

\begin{itemize}
    \item Galaxies from all the three post-merger clusters show spatial distribution implying that the major merging event has resulted in uni-modal $z-$distribution and elongated spatial galaxy redistribution in A3667 \& A3376 but disrupted the low-mass cluster A168. 
    \item We present the first morphology fractions in cluster mergers with radio relics which show comparable fractions of Es and S0s when compared to relaxed clusters. Spiral fraction in relaxed clusters however show significant scatter which is absent in cluster mergers suggesting a uniformly mixed spiral galaxy population with possible extra influx of galaxies from the surrounding cosmic web.
    \item We report independence of colour--magnitude relations on the cluster dynamic state with the post-merger clusters showing red sequence with a near constant scatter similar to relaxed clusters.
    \item Both post-mergers and relaxed clusters show similar blue fractions albeit it with a lot of variation for relaxed clusters. 
    \item We discover that both post-merger and relaxed clusters host considerable populations of red spirals, thus linking their origin to the cumulative pre-merger cluster environment. 
    \item We find that the blue ETG population is characteristic to cluster mergers, and relaxed clusters with known denser large-scale structure connections.

\end{itemize}

In summary, our results thus far point towards newly assembling cluster systems through recent ($\sim$ 1 Gyr) extreme major interaction, to be evolved systems by construction with uniform morphology and colour fractions of member galaxies irrespective of the masses of the newly forming cluster system. By restricting to  clusters which show no obvious interaction (major or minor), the general scatter in the colour and morphology properties of the member galaxies brings to light different global star formation histories for such non-interacting relaxed clusters. We find no clear evidence of post-processing in galaxy colours due to a major cluster merger event, beyond what is expected from our understanding of galaxy evolution in high-density or `cluster environments'. We further propose that the $\sim$ 1 Gyr timescale could be limited to present any observable effects in global colours of galaxies. We also feature merging clusters as unique systems to identify and study filament galaxy populations owing to the extra influx of the cosmic web throughout the merging process. Being the first uniform and spatially matched study to directly compare galaxy populations in very young merging clusters with non-merging ones, our study presents new opportunities and challenges in the field of galaxy evolution in extreme cluster-merger environments.


----------------------------------------- 
----------------------------------------

\begin{acknowledgements}
     KK acknowledges full financial support from ANID through FONDECYT Postdoctrorado Project 3200139, Chile. ACCL thanks the financial support of the National Agency for Research and Development (ANID) / Scholarship Program / DOCTORADO BECAS CHILE/2019-21190049. YJ and ACCL acknowledge financial support from ANID BASAL project No. FB210003 and FONDECYT Iniciaci\'on 2018 No. 11180558. JF and DPM acknowledge financial support from the UNAM-DGAPA-PAPIIT IN110723 grant, Mexico. BV acknowledges the financial contribution from the grant PRIN MIUR 2017 n.20173ML3WW\_001 (PI Cimatti). BV acknowledges support from the INAF Mini Grant 2022 “Tracing filaments through cosmic time”  (PI Vulcani). JPC acknowledges financial support from ANID through FONDECYT Postdoctorado Project 3210709. KK also thanks Arianna Cortesi and Boris Hau{\ss}ler for helpful science discussions. We also thank the Referee for their valuable feedback on this work.

    MGCLS data products were provided by the South African Radio Astronomy Observatory and the MGCLS team and were derived from observations with the MeerKAT radio telescope. The MeerKAT telescope is operated by the South African Radio Astronomy Observatory, which is a facility of the National Research Foundation, an agency of the Department of Science and Innovation. This research has made use of data obtained from the Chandra Data Archive and the Chandra Source Catalog, and software provided by the Chandra X-ray Center (CXC) in the application packages CIAO and Sherpa. 

    This research made use of TOPCAT, an interactive graphical viewer and editor for tabular data \citep{2005ASPC..347...29T}. This research made use of Astropy,\footnote{http://www.astropy.org} a community-developed core Python package for Astronomy \citep{astropy:2018}. This research made use of ds9, a tool for data visualisation supported by the {\it Chandra} X-ray Science Center (CXC) and the High Energy Astrophysics Science Archive Center (HEASARC) with support from the JWST Mission office at the Space Telescope Science Institute for 3D visualisation.

\end{acknowledgements}
\bibliographystyle{aa}
\bibliography{kk_paper6_new}

\begin{appendix} 
\section{Colour--magnitude relations for the control cluster sample and }
\label{ap:cmr-control}

We present colour--magnitude relations for the seven relaxed clusters comprising our control cluster sample, using the recipe described in Section \ref{sec:gcolors} for fitting the red-sequence. The magnitude bins were kept fixed for the fitting for both post-merger and relaxed cluster samples. 

\begin{figure}[t]  
    \includegraphics[width=1\columnwidth]{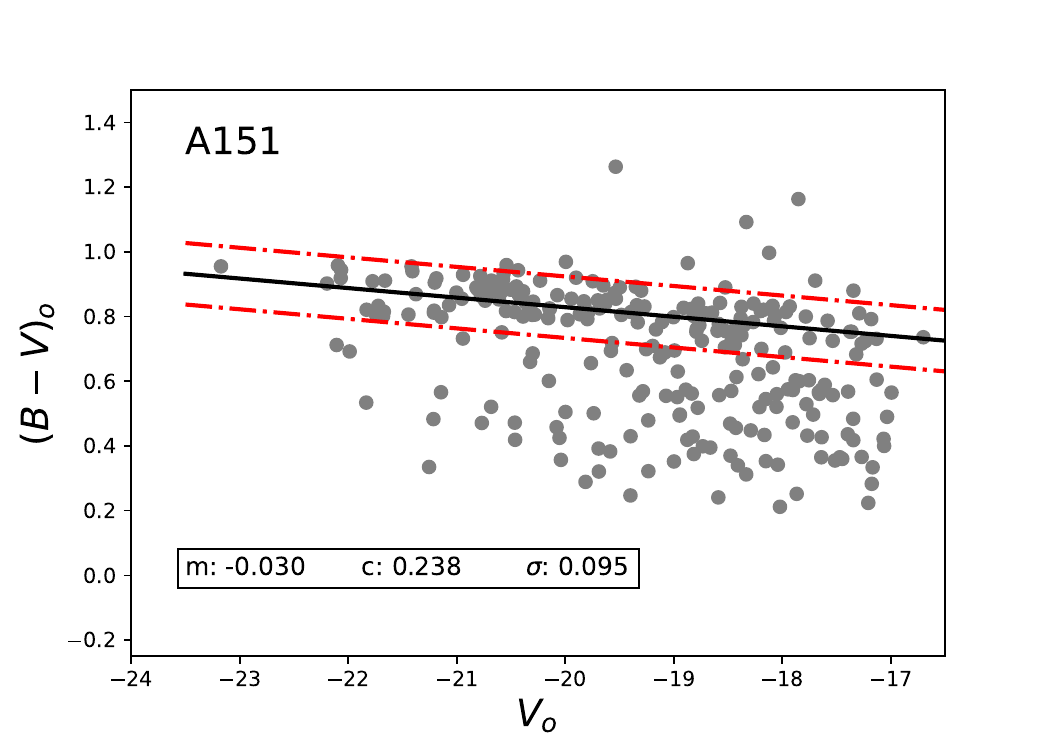}%
    \vspace{-1\baselineskip}
    
    \includegraphics[width=1\columnwidth]{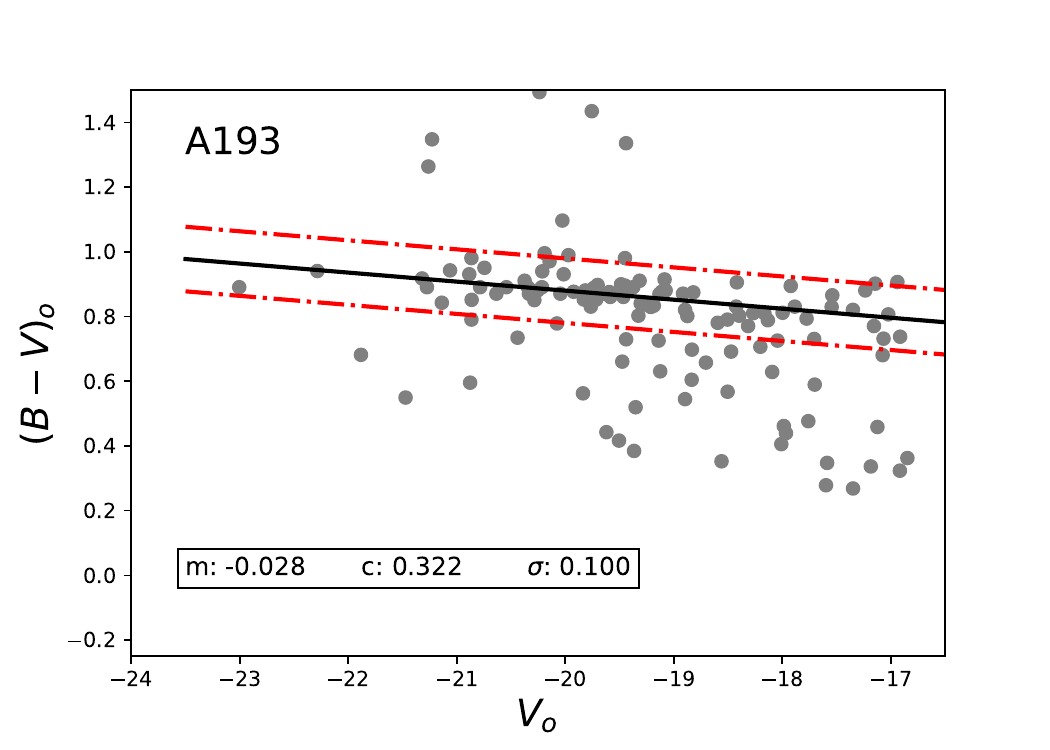}
    \vspace{-1\baselineskip}
    \includegraphics[width=1\columnwidth]{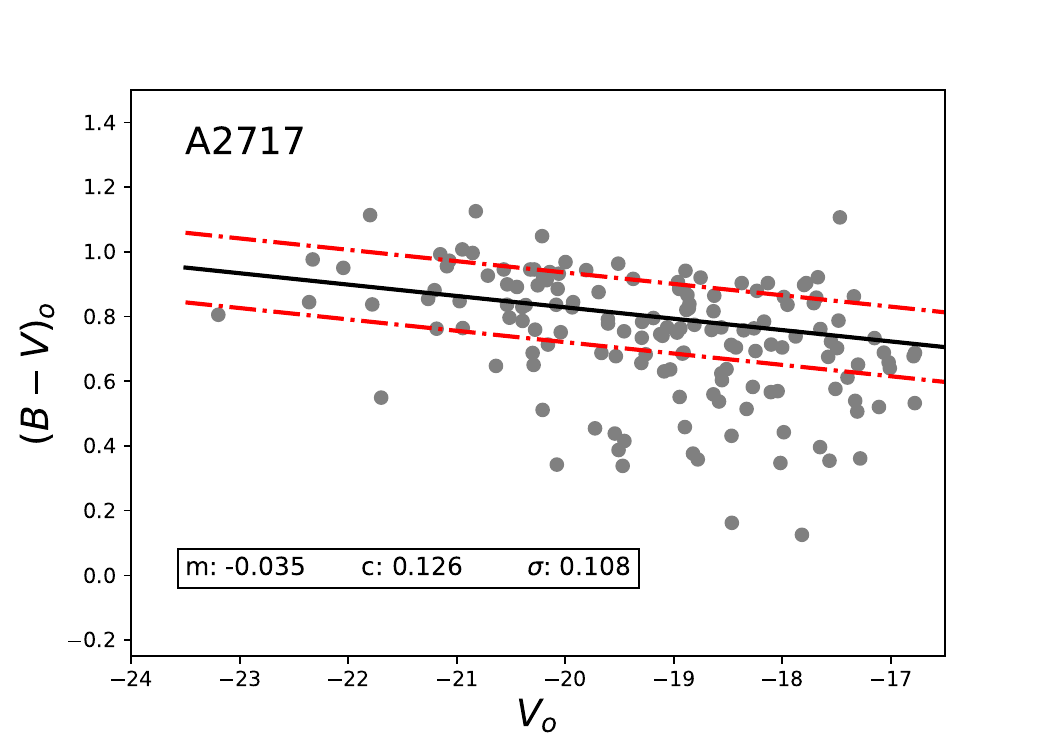}

    \label{fig:app-cmr1}
\end{figure}

\begin{figure}      
    
    \includegraphics[width=1\columnwidth]{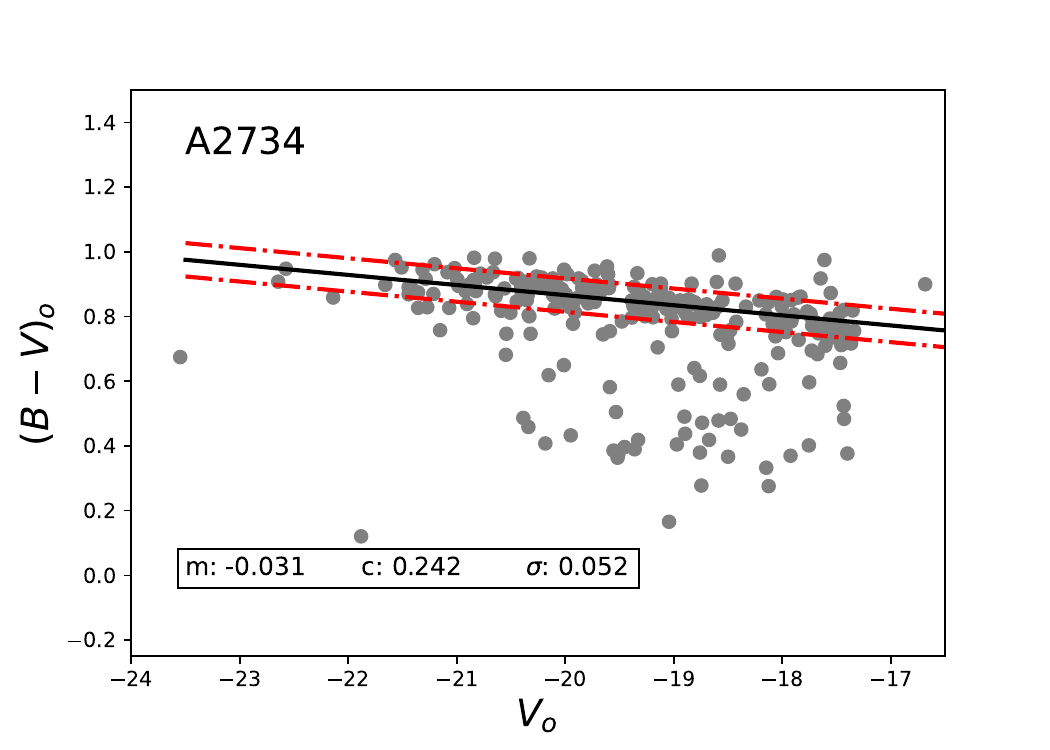}
    \vspace{-1\baselineskip}
    \includegraphics[width=1\columnwidth]{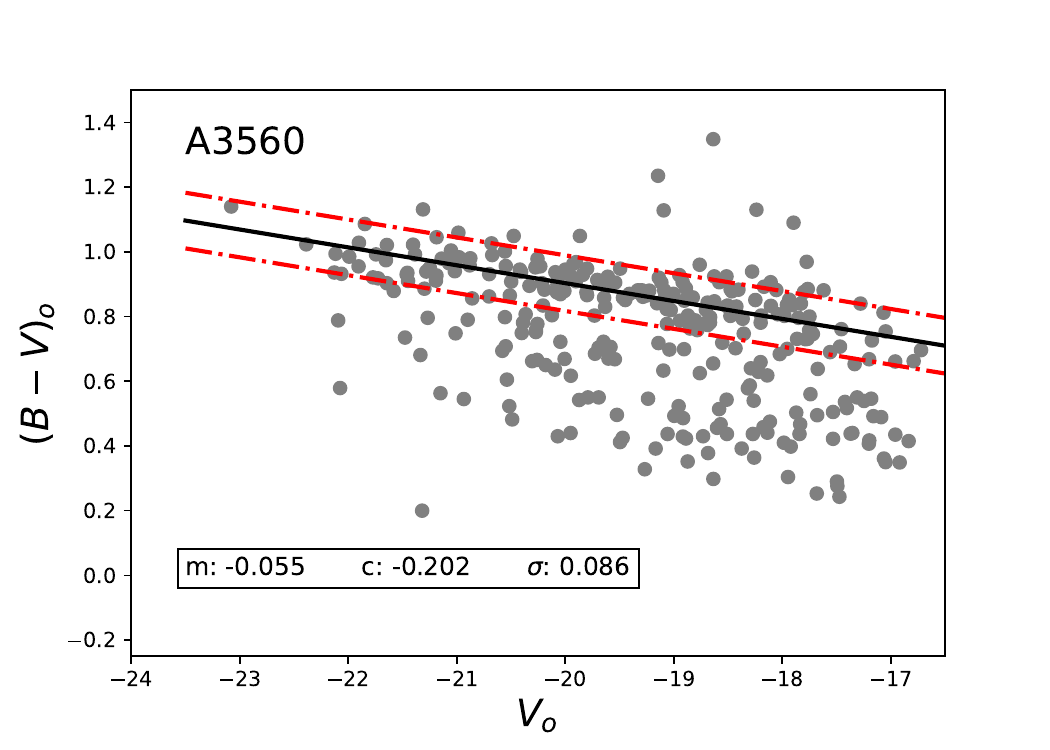}
    \vspace{-1\baselineskip}
    \includegraphics[width=1\columnwidth]{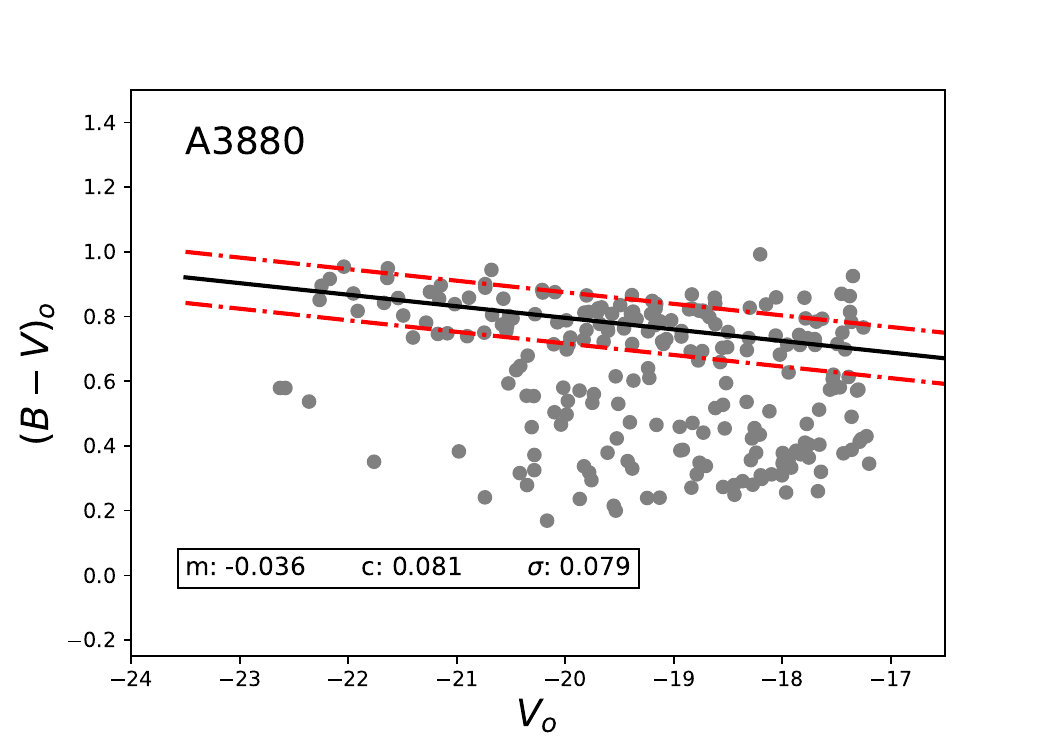}%
    
    \caption{Colour--magnitude relations for control cluster samples. Each panel represents the colour--magnitude relation for a single cluster from our control relaxed cluster sample with the cluster members (grey filled circles) from \citet{vulcani22}. The solid black line indicates the fitted red sequence with the average scatter of the fit ($\sigma$) shown by the red dash-dot lines above and below the fit line. The inset panel gives the details of the fit for each and every cluster. }
\end{figure}

\begin{figure*}[ht]
\centering
  
\includegraphics[width=0.5\linewidth]{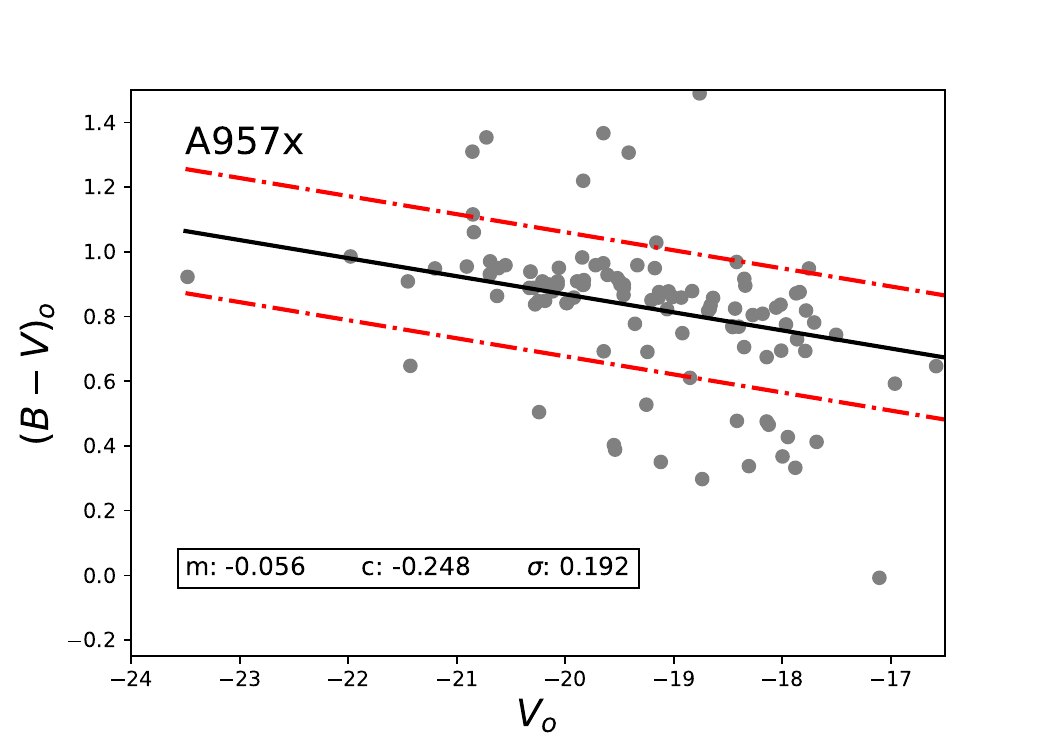}  
\caption{Colour--Magnitude relation for each of the control relaxed clusters (Continued)}
\label{fig:fig}
\end{figure*}
\FloatBarrier
\section{Properties of spiral galaxies in the cluster sample}

\begin{figure}[h]

  \centering
  
  \includegraphics[width=.85\linewidth]{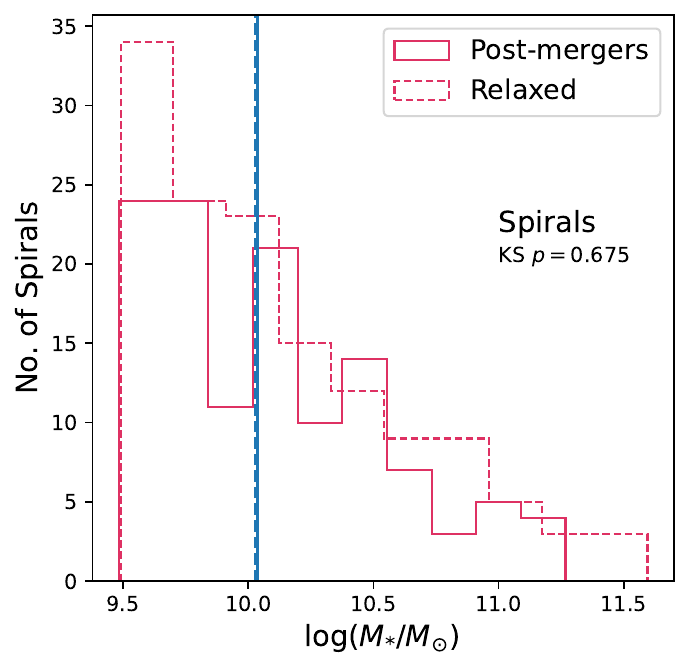}  
  \caption{Mass distribution for spirals in the post-merger and relaxed cluster samples within 0.9$R_{200}$, the aperture we use to compute fractions throughout this work. The inset $p-$value denotes the Kolmogorov-Smirnov test significance under the null hypothesis that the stellar mass distributions from both cluster samples are drawn from the same parent distribution. The vertical solid and dashed lines indicate the median quantities for galaxies in the post-merger and relaxed clusters respectively. }
  \label{fig:a2}
\end{figure} 

\begin{figure}
\vspace{1.7cm}
  \centering

  \includegraphics[width=.85\linewidth]{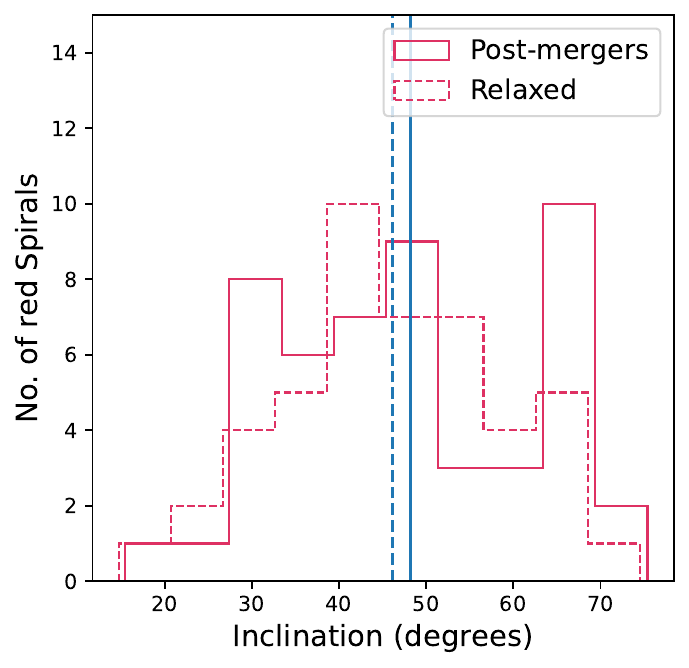}

\caption{Distribution of inclination of red spirals in the post-merger and relaxed cluster samples (Figure \ref{fig:frac_redsp}), computed from the b/a ratios from \citet{gull15}. The vertical solid and dashed lines indicate the median quantities for galaxies in the post-merger and relaxed clusters respectively.}
\label{fig:a3}
\end{figure}

\begin{table*}
\vspace{2\baselineskip}
\centering
\caption{Properties of the WINGS/OmegaWINGS cluster sample}
\begin{tabular}{ c c c c c c c}
\hline

Cluster & $z$ & $\sigma_{\textrm{cl}}$ & $R_{200}$ & log($L_{\rm x}$) &$N_{spec} $& Dynamical\\ 
 & & km s$^{-1}$ & Mpc & erg s$^{-1}$ & & state \\ 

\hline
&\\
A151	& 0.05327 & 771	& 1.670 & 44.00	 &235 & 2 \\
A193	& 0.04852 & 758	& 1.580 & 44.19    & 67 & 2 \\
A2717	& 0.04989 & 470	& 1.170 & 44.00   & 130 & 2 \\
A2734	& 0.06147 & 588 & 1.380 & 44.41  & 215 & 2 \\
A3560	& 0.04917 & 799	& 1.790 & 44.12  & 275 & 3 \\
A3880	& 0.05794 & 514	& 1.200 & 44.27  & 212 & 2 \\
A957x	& 0.04496 & 631	& 1.420 & 43.89  & 77 & 3 \\
\hline
& & & & & & \\
A168	& 0.04518 & 498	& 0.970 & 44.04 & 137 & 5\\
A3376	& 0.04652 & 756	& 1.650 & 44.39  & 223 & 5 \\
A3667	& 0.05528 & 1031& 2.220 & 44.94  & 384&  5\\
\hline
& & & & & & \\
\multicolumn{7}{l}{\small Columns: redshift ($z$), velocity dispersion ($\sigma_{\rm{cl}}$), virial radius ($R_{200}$),}\\
\multicolumn{7}{l}{\small and X-ray luminosity ($L_{\rm x}$), spectroscopically confirmed cluster members ($N_{spec}$),}\\
\multicolumn{7}{l}{\small cluster dynamical state classification from \citetalias{l22} } \\
\end{tabular}
\label{tab:WINGS_clusters}
\end{table*}

\end{appendix}
%
%

\end{document}